\documentclass[preprint,12pt]{article}

\usepackage{amssymb}
\usepackage{arxiv}
\usepackage[T1]{fontenc}
\usepackage{hyperref}
\usepackage{url}
\usepackage{booktabs}
\usepackage{nicefrac}
\usepackage{microtype}
\usepackage{doi}
\usepackage{amsfonts}
\usepackage{graphicx}
\usepackage{epstopdf}
\usepackage{algorithmic}
\usepackage{mathtools}
\usepackage{bm}
\usepackage{tikz}
\usetikzlibrary{arrows,positioning}
\usepackage{pgfplots}
\usepackage{amsopn}
\usepackage{amsmath}
\DeclareMathOperator*{\argmin}{arg\,min}
\usepackage[scientific-notation=true]{siunitx}

\title{A high-order fully Lagrangian particle\\level-set method for dynamic surfaces}

\date{}

\author{ \href{https://orcid.org/0000-0003-2915-8920}{\includegraphics[scale=0.06]{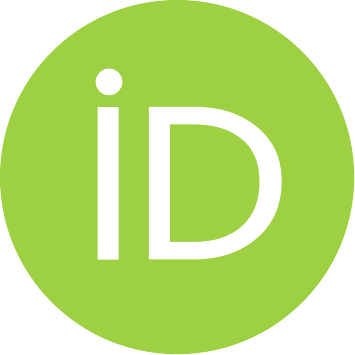}\hspace{1mm}Lennart J.~Schulze}\\
	\texttt{lschulze@mpi-cbg.de}
	\And
	\href{https://orcid.org/0000-0003-4852-2839}{\includegraphics[scale=0.06]{figures/orcid.pdf}\hspace{1mm}Sachin K.~T.~Veettill}\\
	\texttt{sthekke@mpi-cbg.de}
	\And
	\href{https://orcid.org/0000-0003-4414-4340}{\includegraphics[scale=0.06]{figures/orcid.pdf}\hspace{1mm}Ivo F.~Sbalzarini}\thanks{Corresponding author.}\\
	\texttt{sbalzarini@mpi-cbg.de} \\
	\\
	Technische Universit\"at Dresden, Faculty of Computer Science, Dresden, Germany,\\Max Planck Institute of Molecular Cell Biology and Genetics, Dresden, Germany,\\Center for Systems Biology Dresden, Dresden, Germany
}

\renewcommand{\headeright}{}
\renewcommand{\undertitle}{}
\renewcommand{\shorttitle}{A fully Lagrangian particle level-set method}

\hypersetup{
pdftitle={A fully Lagrangian particle level-set method},
pdfauthor={L.J. Schulze, S.K.T. Veettil and I.F. Sbalzarini},
}

\begin{document}
\maketitle

\begin{abstract}
We present a fully Lagrangian particle level-set method based on high-order polynomial regression. This enables closest-point redistancing without requiring a regular Cartesian mesh, relaxing the need for particle-mesh interpolation. Instead, we perform level-set redistancing directly on irregularly distributed particles by polynomial regression in a Newton-Lagrange basis on a set of unisolvent nodes. We demonstrate that the resulting particle closest-point (PCP) redistancing achieves high-order accuracy for 2D and 3D geometries discretized on highly irregular particle distributions and has better robustness against particle distortion than regression in a monomial basis. Further, we show convergence in a classic level-set benchmark case involving ill-conditioned particle distributions, and we present an application to an oscillating droplet simulation in multi-phase flow.
\end{abstract}

\keywords{Level-set methods \and particle methods \and geometric computing \and multi-phase flow}

\section{Introduction}
The numerical representation of non-parametric surfaces is a key part of many spatio-temporal simulations, e.g., in additive manufacturing~\cite{chen2017three}, geology \cite{li2008level}, and biology \cite{sbalzarini2006simulations}. This has motivated research into geometric computing algorithms that can achieve high accuracy for representing non-parametric surfaces at low computational cost. Ideally, the algorithms should also be parallelizable in order to leverage high-performance-computing and GPU resources. 

Due to their geometric expressiveness and parallelizability, level-set methods \cite{osher1988} have emerged as a popular approach to geometric computing for non-parametric surfaces. They have been successfully applied to real-world problems involving complex geometries, such as the growth, division, and reconnection of tumors \cite{zheng2005nonlinear} and the diffusion of membrane proteins on the highly curved endoplasmic reticulum \cite{sbalzarini2006simulations}. In level-set methods, a two-dimensional surface embedded in a three-dimensional space is described implicitly as the (usually zero) level-set of a scalar function over the embedding space. Often, the level-set function is chosen to be the signed distance function (SDF) to the surface. This guarantees that the level-set function is smooth and continuously differentiable near the surface~\cite{sethian2003level}, and it simplifies surface computations as the level-set function value directly represents the shortest distance to the surface. 

Level-set methods also generalize well to dynamically moving and deforming surfaces, as changes in the shape of the surface amount to advecting the level-set function. 
When advecting the level-set function, however, the signed-distance property is in general not conserved. Indeed, for deformation velocity fields that do not describe rigid-body motion, the advection of the level-set function in the embedding space destroys the signed-distance property~\cite{sussman1994level}. This can be avoided by correcting the advection velocity away from the surface, for example by variational penalties~\cite{li2005level,li2010distance} or Lagrange multipliers~\cite{estellers2012efficient}. These approaches, however, are often inaccurate or accumulate advection errors over time.
The most popular approach, therefore, is to recompute the SDF whenever the surface has deformed. This is known as {\em level-set redistancing}, where the SDF in the embedding space is recomputed from the reconstructed current location of the surface. 

Several conceptually different methods for level-set redistancing are available: Sussman et al.~\cite{sussman1994level} performed redistancing by evolving an auxiliary Partial Differential Equation (PDE) in pseudo-time, which has the SDF as a steady-state solution. Their approach achieves high accuracy on regular Cartesian grids and avoids numerical instabilities of the pseudo-time evolution by using higher-order ENO/WENO finite-difference
schemes \cite{liu1994weighted}.
The resulting algorithm, however, is computationally expensive as it amounts to evolving a PDE to steady state in the embedding space with sometimes stringent time-step limitations~\cite{hieber2005lagrangian}. The point of computational efficiency has been addressed by fast marching or sweeping methods \cite{tsitsiklis1995efficient,sethian1996fast,zhao2005fast}, which propagate the level-set values from the surface outward as a moving front. While this is computationally more efficient, it is limited to lower-order finite-difference schemes for which the resulting algebraic equations can be analytically solved, and it creates data dependencies that hamper parallelization \cite{kim2001AnON}. 
A third approach therefore aims to directly compute the distance to the surface independently for all points in the embedding space by finding for each query point the closest point on the surface \cite{mauch2000fast}. This {\em closest-point} (CP) transform has mainly found application in the numerical solution of surface PDEs~\cite{ruuth2008simple,marz2012calculus} and has since been extended to level-set resdistancing using higher-order polynomial regression~\cite{saye2014}. The resulting method achieves high orders of accuracy, is computationally efficient, and parallelizable. While originally formulated for regular Cartesian grids, it has recently also been demonstrated on unstructured grids \cite{henneaux2021high} and triangulations~\cite{ngo2022high}, confirming the versatility of the CP approach.

Despite the efficiency and versatility of CP redistancing, however, algorithms that achieve high orders of convergence are so far limited to connected meshes and discretize level-set advection in an Eulerian frame of reference by evolving the level-set function values at the mesh nodes. In such an Eulerian mesh-based approach, fulfilling conservation laws becomes nontrivial, and the numerical stability and adaptivity of the overall framework is limited by the CFL condition. Both limitations can be relaxed when discretizing level-set advection in a Lagrangian frame of reference, where the discretization points move with the local advection velocity and preserve their level-set function values. Lagrangian level-set methods have been shown to be highly geometry-adaptive with excellent numerical stability~\cite{hieber2005lagrangian} while maintaining the general conservation properties~\cite{monaghan2005smoothed} of Lagrangian particle methods.

In Lagrangian particle methods, however, particle distributions become increasingly irregular as particles move with the advection velocity. This hampers level-set redistancing, for which so far only first-order methods exist that aim to regularize the level-set function by renormalization~\cite{engquist2005discretization,cottet2006level}. Previous higher-order approaches exclusively operate on grids, which requires interpolation of the level-set function values from the Lagrangian particles to the grid nodes before redistancing~\cite{enright2002hybrid,hieber2005lagrangian}. This introduces additional computational cost and interpolation errors. While high-order particle-to-mesh interpolation schemes exist~\cite{cottet2014high}, they are based on conservation laws for the moments of the represented function, which do not apply to level-set functions. While again renormalization approaches have been proposed to address the problem~\cite{bergdorf2010lagrangian}, their convergence is limited to linear order.

Here, we present a fully Lagrangian particle CP level-set redistancing method that achieves higher-order convergence without requiring  interpolation to a structured or unstructured intermediate mesh. The method directly operates on Lagrangian particles, maintaining their conservation properties and stability, while simplifying level-set advection for dynamic surfaces by inheriting the accuracy, computational efficiency, and parallelizability of CP redistancing. In the present method, particles in a narrow-band around the surface carry and advect the level-set function values. After advection, the SDF is recomputed directly on the irregularly distributed particles by finding their respective closest points on the surface. We do this using high-order polynomial regression in a Newton-Lagrange basis on unisolvent nodes. The analytical form of the local regression polynomials enables straightforward computation of derivative geometric quantities, such as surface normals and local curvatures. The regression nodes are a suitably chosen subset of particles in the narrow-band around the surface. We show that a clever choice of local unisolvent nodes maintains high-order convergence even on highly irregular particle distributions in the narrow-band. We discuss accuracy, stability, and performance of the method on benchmark geometries with analytically known SDF, a vortex flow problem, and a multi-phase hydrodynamics application. In the latter, we compare the present approach with a Smoothed Particle Hydrodynamics (SPH) method~\cite{morris2000simulating}.

\section{Level-set method}
Level-set methods describe an evolving surface $\Gamma_t$ implicitly as the zero level-set of a scalar function $\phi$:
\begin{equation}
    \Gamma_t=\{\mathbf{x}\,:\,\phi(\mathbf{x},t)=0\},
\end{equation}
with $\mathbf{x}\in\mathbb{R}^{n_d}$ 
being the coordinates in $n_d$-dimensional space, and $t$ being the time. Due to favourable properties in accuracy and volume conservation, as well as simplified computations of surface-geometric quantities such as normals and curvatures, a popular choice for the level-set function is a signed distance towards the surface:
\begin{equation}\label{eq:sdf}
    \phi(\mathbf{x})=\pm\|\mathrm{\mathbf{cp}}(\mathbf{x})-\mathbf{x}\|_2,
\end{equation}
in which $\mathrm{\mathbf{cp}}(\mathbf{x})$ is the closest-point function that yields the closest point of a given location on the surface, measured in $L_2$-distance. The sign of the level-set function \eqref{eq:sdf} is chosen as positive if outside, and negative if inside of a closed surface. A natural property of the SDF \eqref{eq:sdf} is that its gradient has unit length,
\begin{equation}\label{eq:unitgradient}
    \|\nabla\phi(\mathbf{x})\|_2 = 1.
\end{equation}
Level-set functions also allow for a computation of derivative fields associated with surfaces, such as the surface normal
\begin{equation}\label{eq:normal}
    \mathbf{n}=\frac{\nabla\phi(\mathbf{x})}{\|\nabla\phi(\mathbf{x})\|_2}
\end{equation}
and the local mean curvature
\begin{equation}\label{eq:curvature}
    \kappa=\nabla\cdot\mathbf{n}
\end{equation}
if $n_d=2$, and $\kappa=\frac{1}{2}\nabla\cdot\mathbf{n}$ if $n_d=3$. 

The surface $\Gamma_t$ can move and deform with velocity $\mathbf{u}\left(\mathbf{x}, t\right)$ over time $t$. After any movement, material points lying on the surface remain on the surface:
\begin{equation}\label{eq:evolutionLag}
    \frac{\text{D}\phi}{\text{D}t}=0.
\end{equation}
Eq. \eqref{eq:evolutionLag} is formulated using the material derivative $\frac{\text{D}(\cdot)}{\text{D}t}=\frac{\partial(\cdot)}{\partial t} + \mathbf{u}\cdot\nabla(\cdot)$ and generally only holds for material points on the surface, i.e. on the zero level-set. Material points surrounding the surface can either approach to or recede from the surface, which is not accounted for by Eq.~\eqref{eq:evolutionLag}, as the velocity field $\mathbf{u}$ is embedded in the 3D space. If changes in distance towards the surface are not taken into account, the level-set function ceases to be the SDF and the property \eqref{eq:unitgradient} is lost, hampering the computation of surface normals and local curvatures. Therefore, level-set redistancing is used to restore the SDF property.

\section{Redistancing on irregular particle distributions}
We represent the surface $\Gamma_t$ using a finite set of $n_p$ discrete Lagrangian particles $P_i$, $i\in\{1,\ldots ,n_p\}$. These particles store and advect the level-set function values $\phi_i$ in addition to their position. At the beginning of a simulation, the particles are usually seeded uniformly or uniformly at random with inter-particle spacing $h$ in a tubular neighborhood of diameter $w$ around the surface, which we call {\em narrow-band} in accordance with the usual level-set terminology. Then, the particles move with the deformation velocity $\mathbf{u}$ of the surface, and the particle distribution can become arbitrary.

This Lagrangian formulation offers a number of benefits: rigid-body movements of the surface are simulated exactly~\cite{monaghan2005smoothed}, aside from time integration errors, and the method has better time-integration stability as the Lagrangian CFL number is larger than the Eulerian CFL number \cite{bergdorf2006lagrangian}, and no grid data structure needs to be allocated and maintained.

\begin{figure}[ht]
    \centering
    \input{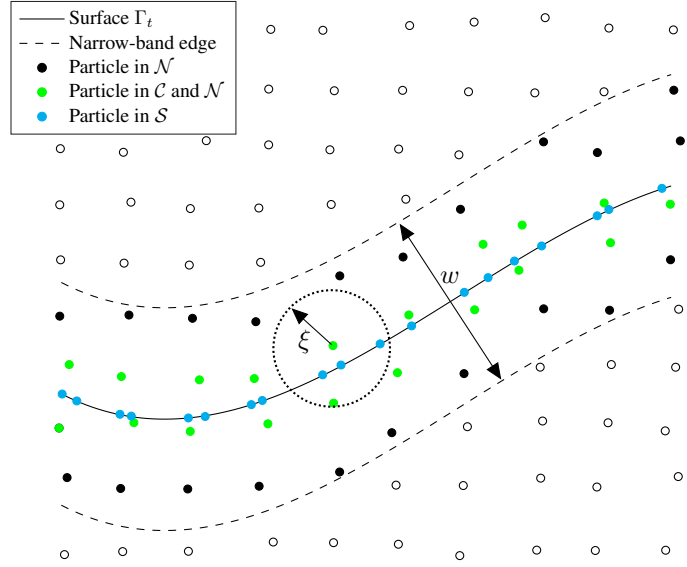}
    \caption{Discretized 2D domain containing surface.}
    \label{fig:narrowband}
\end{figure}

We extend the CP redistancing proposed by Saye \cite{saye2014} to irregularly distributed particles in a narrow-band of width $w$ around the surface, as illustrated in Fig.~\ref{fig:narrowband}. The set containing all particles of the narrow-band is $\mathcal{N} = \{P_i \,:\, |\phi(\mathbf{x}_i)| < w/2\}$.

Before redistancing, an additional set of {\em sample particles} $\mathcal{S}$ that lie precisely on the surface is created. For this, particles immediately adjacent to the surface are used as starting points (set $\mathcal{C}$ in Fig.~\ref{fig:narrowband}). Using local polynomial regression over neighborhoods of such near-surface particles, the level-set function $\phi (\mathbf{x})$ is locally approximated. The  polynomial approximation is then used to project the location of the particle onto the zero level-set, yielding the corresponding sample particle. 

For redistancing, the following steps are then computed for all particles in $\mathcal{N}$:
First, we find the closest sample particle using a cell-list acceleration data structure \cite{quentrec1973new}, which serves as an initial guess for solving a constrained optimization problem. This problem minimizes the distance between the query particle and another point under the constraint that the polynomially approximated level-set function at the location of the other point is zero. This is optimized over the same local regression polynomial used to generate the sample particles and yields the closest point on the zero level-set of the local regression polynomial. The distance between the so-computed closest point and the query particle is the corrected level-set value that restores the SDF. 

This procedure crucially hinges on the numerical properties of the polynomial regression scheme used. In the spirit of particle methods, we perform polynomial regression in a local radial neighborhood of a particle. The radius $r_c$ of this neighborhood needs to include at least as many particles as are required to determine the coefficients of the regression polynomial for a given polynomial degree. Further, the spatial arrangement of the particles within the regression neighborhood cannot be dependent in a way that renders the Vandermonde matrix of the regression problem singular. In the next section, we therefore pay particular attention to how polynomial regression is done here.

\subsection{Local polynomial approximation of the level-set function}
The sample particles provide a coarse estimate of the surface and are obtained by first finding a set $\mathcal{C}\subset\mathcal{N}$ containing particles in the narrow-band that are close to the surface. A particle is in $\mathcal{C}$ if it has another particle with opposite sign of the level-set function value within a certain radius $\xi$. 
This radius is a parameter of the method. Larger thresholds create more sample particles, thus improving the sampling of the surface, but causing higher computational cost. In this paper, we use a threshold radius of $\xi=1.5h$ throughout. Then, $\mathcal{C}$ is a small subset of $\mathcal{N}$.

For each particle $P_i\in\mathcal{C}$, we approximate the level-set function as a continuous polynomial obtained by local least-squares regression. 
As we show in Sec.~\ref{sec:remeshingredistancing}, this is simpler and more accurate than directly using a particle-function approximation of the level-set function. Using $n_c$ monomials $M_k\left(\mathbf{x}\right)$, the local approximation of the level-set function reads:
\begin{equation}\label{eq:poly_expansion}
    \phi(\mathbf{x})\approx p_i(\mathbf{x})=\sum_{k=0}^{n_c}c_k M_k(\mathbf{x})\, .
\end{equation}
The choice of the number and type of monomials can be made depending on the requirements of the application. 

We determine the coefficients $c_k$ of $p_i(\mathbf{x})$ by iterating over the neighborhood of a particle $P_i\in\mathcal{C}$ that contains $n_n$ particles including $P_i$ itself. In this neighborhood, we assemble the regression matrix $\mathbf{A}\in\mathbb{R}^{n_n\times n_c}$, the unknown coefficient vector $\mathbf{c}\in\mathbb{R}^{n_c}$, and the right-hand side vector containing the level-set function values $\bm{\phi}\in\mathbb{R}^{n_n}$. Then, the linear system of equations
\begin{equation}\label{eq:LSE}
    \mathbf{A}\mathbf{c}=\boldmath\phi
\end{equation}
is solved using orthogonal decomposition. If $n_n=n_c$ and $\mathbf{A}$ has full rank, the polynomial $p_i(\mathbf{x})$ is the interpolation polynomial. If $n_n>n_c$, $p_i(\mathbf{x})$ is the least-squares solution for the polynomial regression problem.

The condition number of the regression matrix depends on the distribution of particles and the choice of basis $M_k(\mathbf{x})$. We do not have control on the former, but we can choose the polynomial basis $M_{k}(\mathbf{x})$. As a basis, we use the Lagrange polynomials $L_k(\mathbf{x})$ determined on a Chebyshev-Lobatto grid $G\subseteq [-1,1]^{n_d}$ in the regression neighborhood, i.e., $L_k(\mathbf{q}_l) = \delta_{k,l}$, $\mathbf{q}_l \in G$, where $\delta_{k,l}$ is the Kronecker delta \cite{hecht2020multivariate}. Therefore, the points at which the polynomial basis is computed (i.e., the nodes of a local Chebyshev-Lobatto grid) in general differ from the points $p_i(\mathbf{x})$ at which the regression polynomial is evaluated. We use basis polynomials of total degree ($\ell_1$-degree) \cite{hecht2020multivariate}. 
The regression matrix is constructed by evaluating the Newton form of the Lagrange basis polynomials on the regression nodes. This choice of basis has been shown to effectively regularize regression over randomly distributed points for a large class of analytic functions~\cite{veettil2022multivariate}. We refer to this regression approach as {\em minter} regression.

Since Eq.~\eqref{eq:evolutionLag} holds on the surface after any movement, solving Eq.~\eqref{eq:LSE} for each particle in $\mathcal{C}$ yields a polynomial representation of the surface as the zero level-set of the local regression polynomials $p_i(\mathbf{x})$. Using this representation in the proximity of the ``center'' particle $P_i\in\mathcal{C}$, the locations of the sample particles $\mathcal{S}$ can be determined. This is done by iterative projection onto the zero level-set of $p_i(\mathbf{x})$, using $P_i\in\mathcal{C}$ as a starting point and iterating
\begin{equation}\label{eq:surfaceprojections}
    \mathbf{x}^{k+1} = \mathbf{x}^k-p_i(\mathbf{x}^k)\frac{\nabla p_i(\mathbf{x}^k)}{\|\nabla p_i(\mathbf{x}^k)\|_2^2}, \qquad k=0,1,2,\ldots
\end{equation}
The iteration is stopped as soon as $|p_i\left(\mathbf{x}^k\right)|<\varepsilon$ for a user-defined tolerance $\varepsilon$. 

Doing so for all particles in $\mathcal{C}$ completes the sample particle set $\mathcal{S}$, as illustrated in Fig.~\ref{fig:narrowband}. The sample particles provide starting points for the subsequent search for the closest point of any query point (also between particles). They further act to store the regression polynomials and therefore the local geometry of the surface. The resolution with which $\mathcal{S}$ samples the surface results from the particle distribution around the surface: Each particle in $\mathcal{C}$, from both sides of the surface, creates one sample particle.
For example, a straight 1D surface of length $l$ embedded in 2D space, in which the domain has been discretized using an inter-particle spacing of $h$, would contain $2l/h$ sample particles.

\subsection{Finding the closest point on the surface for a given particle}

From the sample particles $\mathcal{S}$ and the associated polynomials, the distance of any given location towards the surface can be computed. The query point may also lie between particles or outside of the narrow-band. In level-set redistancing, however, typically the query points are the particles $\mathcal{N}$ within the narrow-band.

For each query particle $P_q\in\mathcal{N}$ at position $\mathbf{x}_q$, we look for the closest sample particle $P_s\in\mathcal{S}$, located at position $\mathbf{x}_s$. In a well-sampled surface, we know that the closest point of $P_q$ lies in a neighborhood of its closest sample particle. Hence, the zero level-set of the regression polynomial $p_s$ is used as a local approximation of the surface. To find the closest point $\mathbf{x}$ of $P_q$ on the approximated surface, we solve the constrained optimization problem
\begin{equation}
    \argmin_\mathbf{x} \frac{1}{2}\|\mathbf{x}-\mathbf{x}_q\|_2^2\, , \quad\text{ s.t. } p_s(\mathbf{x})=0\, ,
\end{equation}
minimizing the distance under the constraint that the solution lies on the zero level-set of the regression polynomial. We reformulate this problem using a Lagrange multiplier $\lambda$ and the associated Lagrangian
\begin{equation}\label{eq:lagrangian}
    \mathcal{L}(\mathbf{x},\lambda)=\frac{1}{2}\|\mathbf{x}-\mathbf{x}_q\|_2^2+\lambda p_s(\mathbf{x})\, .
\end{equation}
Stationary points of the Lagrangian fulfill 
\begin{equation}\label{eq:gradientlagrangian}
\nabla_{\mathbf{x},\lambda}\mathcal{L}=\left(
\begin{matrix}
\mathbf{x}-\mathbf{x}_q+\lambda\nabla p_s(\mathbf{x})\\
p_s(\mathbf{x})
\end{matrix}
\right)=\mathbf{0}
\end{equation}
and are found using the Newton method. The subscripts $\mathbf{x}$ and $\lambda$ indicate differential operators with respect to both variables. As an initial guess $\mathbf{x}^0$ for the iterative Newton method, we use the location of the closest sample particle: $\mathbf{x}^0=\mathbf{x}_s$. The initial Lagrange multiplier is  $\lambda^0=\left(\mathbf{x}_q-\mathbf{x}^0\right)\cdot\nabla p_s(\mathbf{x}^0)/\|\nabla p_s(\mathbf{x}^0)\|_2^2$. Subsequently, we iterate
\begin{equation}\label{eq:newton}
    \left(\begin{matrix}
    \mathbf{x}^{k+1}\\
    \lambda^{k+1}
    \end{matrix}\right)=\left(\begin{matrix}
    \mathbf{x}^k\\
    \lambda^k
    \end{matrix}\right)-\left(\text{H}_{\mathbf{x},\lambda}\mathcal{L}(\mathbf{x}^k,\lambda^k)\right)^{-1}\nabla_{\mathbf{x},\lambda}\mathcal{L}(\mathbf{x}^k,\lambda^k),
\end{equation}
where H is the Hessian, computed as 
\begin{equation}
    \text{H}_{\mathbf{x},\lambda}\mathcal{L}(\mathbf{x},\lambda)=\left(
    \begin{matrix}
    I+\lambda \text{H} p_s(\mathbf{x}) & \nabla p_s(\mathbf{x})\\
    \nabla p_s(\mathbf{x})^\top & 0
    \end{matrix}\right).
\end{equation}
We perform the iterations in Eq.~\eqref{eq:newton} until the $L_2$-norm of the gradient in Eq.~\eqref{eq:gradientlagrangian} falls below the tolerance $\varepsilon$, or a maximum number of iterations $k_{\mathrm{max}}$ is reached.

If the Newton iterations stray out of the support neighborhood of the sample point $P_s$, this can be detected and a new iteration can be started from the corresponding neighboring sample point. In all benchmarks presented in this paper, however, we did not encounter such a case.

Following the definition in Eq.~\eqref{eq:sdf}, we use the resulting approximation of the closest point $\mathbf{x}\approx\mathrm{\mathbf{cp}}(\mathbf{x}_q)$ to update the level-set function value of the query particle $P_q$:
\begin{equation}\label{eq:sdfupdate}
    \phi_q=\text{sgn}(\phi_q)\|\mathbf{x}-\mathbf{x}_q\|_2\, .
\end{equation}
Here, sgn denotes the sign function that ensures that the sign of the new SDF value is the same as the sign of the old level-set function value. 

\subsection{Derivative surface quantities}

From the closest point of a given query particle and the polynomial approximation of the level-set function in the vicinity of the closest point, it is straightforward to compute derivative surface quantities, such as the surface normal and the local mean curvature at the closest point.

We compute the surface normal at the closest point of a query particle $P_q$ as
\begin{equation}
    \mathbf{n}_q = \frac{\nabla p_s\left(\mathbf{cp}(\mathbf{x}_q)\right)}{\|\nabla p_s\left(\mathbf{cp}(\mathbf{x}_q)\right)\|_2}\, ,
\end{equation}
and the local mean curvature as 
\begin{equation}\label{eq:discretemeancurvature}
    \kappa_q=\nabla\cdot\mathbf{n}_q
\end{equation}
in 2D simulation embedding domains and $\kappa_q=\frac{1}{2}\nabla\cdot\mathbf{n}_q$ in 3D. Note that as during the previously outlined redistancing method, the gradient and Hessian of the polynomial are known analytically. Evaluating the derivative at the closest point on the surface also yields a constant normal extension of surface normals and curvature values into the narrow-band.

\subsection{Narrow-band updates}
During level-set advection, particles may enter or exit the narrow-band of width $w$. We assign particles to the narrow-band set $\mathcal{N}$ or its complement $\mathcal{N}^c$ depending on their level-set function value:
\begin{align}
    P_i\in
    \begin{cases}
    \mathcal{N},&\text{ if }|\phi_i|\leq \frac{w}{2}\\
    \mathcal{N}^c,&\text{ if }|\phi_i|>\frac{w}{2}\, .
    \end{cases}
\end{align}
Keeping track of particles leaving the narrow-band is trivial, since a particle in $\mathcal{N}$ moving outside the narrow-band will be redistanced with a SDF value $>\frac{w}{2}$ and is removed from the set $\mathcal{N}$. 
New particles entering the narrow-band, however, cannot be detected based on their level-set function value, as they have not yet been redistanced. We therefore periodically redistance with an enlarged band width, which we refer to as {\em skin} in analogy to Verlet lists~\cite{verlet1967computer}. The thickness of the skin and the frequency with which it is used both depend on the dynamics of the simulation. At the beginning of a simulation, the skin should cover all particles that could possibly enter the narrow-band during the course of the entire simulation. The frequency of skin redistancing should be chosen such that a particle entering the narrow-band is detected as a narrow-band particle before its geometric information is required in the regression vicinity of the surface.

\begin{figure}[ht]
    \centering
    \includegraphics[scale=0.8]{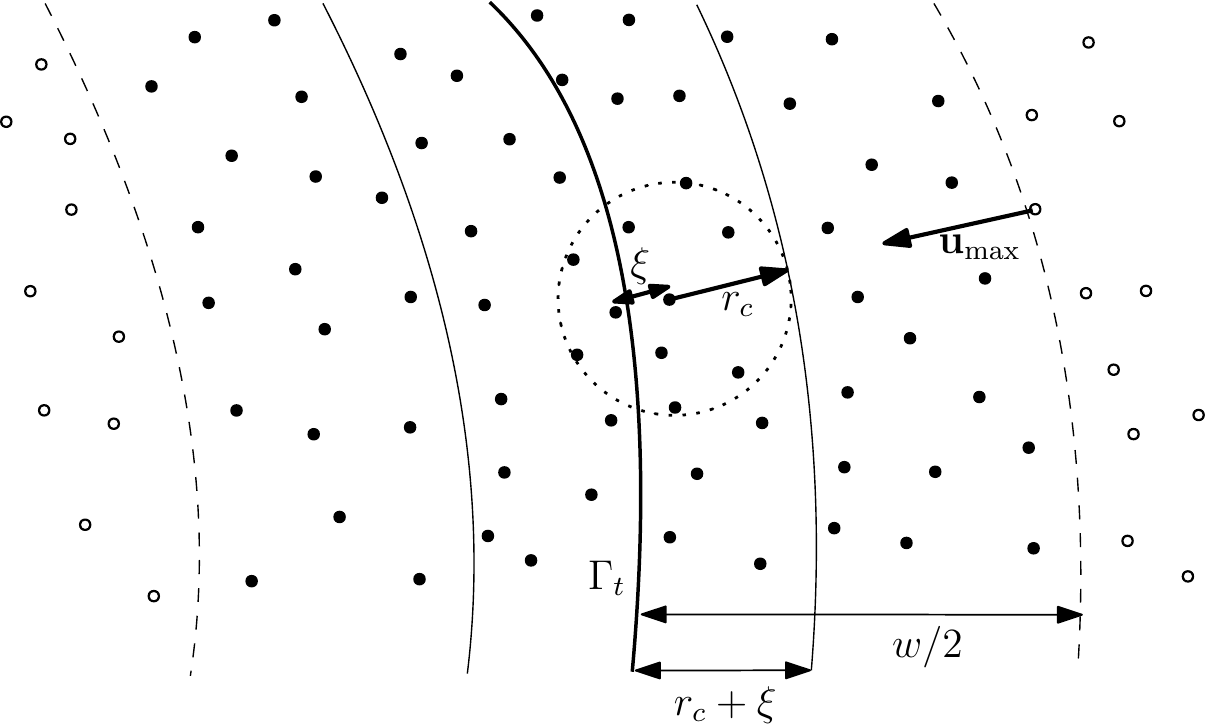}
    \caption{Continuum surface model inside the narrow-band. Solid dots denote particles $\mathcal{N}$ in the narrow-band of width $w$, and hollow dots denote particles in the skin outside the narrow-band. Dashed lines represent the narrow-band borders, and the solid lines surrounding the surface $\Gamma_t$ (bold solid line) delimit the space within which particles contribute to the local regression polynomials (light-gray lines) approximating the surface within a neighborhood radius $r_c$.}
    \label{fig:continuumsurface}
\end{figure}

To determine the frequency of skin redistancing, we consider the minimum time a particle requires to travel $w/2 - (r_c+\xi)$, since $r_c+\xi$ is the maximum distance from the surface below which a particle can contribute to the local regression polynomials, as highlighted in Fig.~\ref{fig:continuumsurface}. With the maximum advection velocity $\mathbf{u}_\mathrm{max}$ anticipated in a simulation, this requires a time of  $\frac{w/2-(r_c+\xi)}{\|\mathbf{u}_\mathrm{max}\|_2}$. Hence, the frequency of skin  redistancing is
\begin{equation}\label{eq:skinfreq}
    f_\mathrm{redist, skin}\geq\frac{\|\mathbf{u}_\mathrm{max}\|_2}{w/2-(r_c+\xi)}
\end{equation}
to ensure detecting any particle relevant to surface dynamics.

\section{Results}
We implement the above {\em Particle Closest-Point} (PCP) method in the scalable scientific computing framework OpenFPM~\cite{incardona2019openfpm} using the minter package \cite{minter} and the Eigen library \cite{eigenweb} for regression. 
We characterize the accuracy and efficiency of the method in benchmarks ranging from basic geometries with known analytical SDF over standard dynamic-surface test cases to a multi-phase flow problem with interfacial effects.

\subsection{Basic geometries}
In his work on mesh-based closest-point redistancing, Saye \cite{saye2014} presents results for elementary geometries such as 2D ellipses and 3D ellipsoids. To demonstrate the utility of a method tailored to data on irregularly distributed particles, we first highlight the problems of the apparent alternative of interpolating particle data to mesh nodes and subsequently applying a mesh-based method. Then, we compare it with the proposed PCP method directly on the particles.

Irregular particle distributions for the benchmarks are obtained by randomly shifting the nodes $\mathbf{m}_{ij}$ of a regular Cartesian mesh with spacing $h$:
\begin{equation}
    \mathbf{x}_p = \mathbf{m}_{ij} + \mathbf{X}\, ,
\end{equation}
with random shifts 
\begin{equation}
    X_d = \alpha h\mu\, ,
\end{equation}
for $d=\{1, \ldots , n_d\}$. The pseudo-random variables $\mu\sim \mathcal{U}[-1, 1]$ are {\it i.i.d.}~from the uniform distribution over the interval $[-1,1]$. The shift amplitude $\alpha$ is always chosen $<0.5$ to ensure that no two particles coincide.

\subsubsection{Remeshing followed by closest-point redistancing}\label{sec:remeshingredistancing}
As a baseline, we first characterize the classic approach of interpolating the particle values to a grid, followed by CP redistancing \cite{saye2014}. For this, we consider an ellipse in the 2D domain $[-1,1]\times[-1,1]$ discretized by particles with spacing $h$ and shift amplitude $\alpha=0.3$ covering the entire domain with periodic boundary conditions. The level-set values at the particles are initialized to
\begin{equation}\label{eq:ellipseeq}
    \phi\left(x,y\right)=1-\sqrt{\frac{x^2}{A^2}+\frac{y^2}{B^2}}\, ,
\end{equation}
for an ellipse with semi-major axis $A=0.75$ and semi-minor axis $B=0.5$. The zero level-set of Eq.~\eqref{eq:ellipseeq} coincides with the zero level-set of the SDF of the ellipse. Away from the surface, however, this is not a SDF. We compute a SDF approximation by remeshing followed by mesh-based CP redistancing according to Saye \cite{saye2014}. 

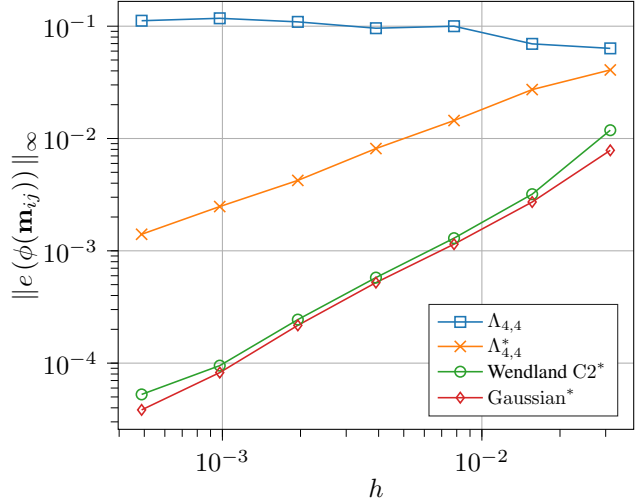
\begin{figure}[ht]
    \centering
    \begin{tikzpicture}

\definecolor{crimson2143940}{RGB}{214,39,40}
\definecolor{darkgray176}{RGB}{176,176,176}
\definecolor{darkorange25512714}{RGB}{255,127,14}
\definecolor{forestgreen4416044}{RGB}{44,160,44}
\definecolor{mediumpurple148103189}{RGB}{148,103,189}
\definecolor{steelblue31119180}{RGB}{31,119,180}

\begin{axis}[
log basis x={10},
log basis y={10},
tick align=outside,
tick pos=left,
x grid style={darkgray176},
xlabel={$h$},
xmajorgrids,
xmin=0.000396607615408318, xmax=0.0384732629170286,
xmode=log,
xtick style={color=black},
y grid style={darkgray176},
ylabel={$\|e\left(\phi(\mathbf{m}_{ij})\right)\|_\infty$},
ymajorgrids,
ymin=2.57386071800733e-05, ymax=0.166265200601574,
ymode=log,
ytick style={color=black},
legend style={legend cell align=left, align=left, draw=white!15!black, nodes={scale=0.75, transform shape}, mark options={scale=1.1}},
legend pos=south east
]
\addplot [semithick, steelblue31119180, mark=square, mark size=2, mark options={solid}]
table {
0.03125 0.063455
0.015625 0.069609
0.0078125 0.099967
0.00390625 0.095931
0.001953125 0.109253
0.0009765625 0.117577
0.00048828125 0.111838
};
\addlegendentry{$\Lambda_{4,4}$}

\addplot [semithick, darkorange25512714, mark=x, mark size=3, mark options={solid}]
table {
0.03125 0.040766
0.015625 0.027260
0.0078125 0.014457
0.00390625 0.008137
0.001953125 0.004236
0.0009765625 0.002478
0.00048828125 0.001404
};
\addlegendentry{$\Lambda_{4,4}^*$}
\addplot [semithick, forestgreen4416044, , mark=o, mark size=2, mark options={solid}]
table {
0.03125 0.0118529
0.015625 0.0031936
0.0078125 0.00129607
0.00390625 0.000577605
0.001953125 0.000244107
0.0009765625 9.50215e-05
0.00048828125 5.26641e-05
};
\addlegendentry{Wendland $\mathrm{C2}^*$}
\addplot [semithick, crimson2143940, mark=diamond, mark size=2, mark options={solid}]
table {
0.03125 0.00784859
0.015625 0.00272034
0.0078125 0.00114637
0.00390625 0.000521113
0.001953125 0.000216951
0.0009765625 8.24069e-05
0.00048828125 3.8351e-05
};
\addlegendentry{$\mathrm{Gaussian}^*$}
\end{axis}

\end{tikzpicture}
    \caption{Convergence of different particle-mesh interpolation schemes (symbols, inset legend) applied to the ellipse Eq.~\eqref{eq:ellipseeq} discretized on irregularly ($\alpha=0.3$) distributed particles. The star in the legend entries indicates that the respective remeshing formulation is renormalized by the particle volumes.}
    \label{fig:p2mconvergence}
\end{figure}

We compare a variety of remeshing schemes as described in Appendix \ref{sec:appendixkernels}. We compute mesh-node values using the $\Lambda_{4,4}$ kernel in both the basic formulation \eqref{eq:lambda44interpol} and the renormalized formulation \eqref{eq:lambda44renormalizedinterpol}, as well as using the renormalized Wendland C2 and Gaussian kernel functions \eqref{eq:particlereprenormalized}. We compare the remeshing results with the analytically known exact values $\phi_{\text{exact}}$ obtained by evaluating Eq.~\eqref{eq:ellipseeq} at the mesh nodes.
The absolute remeshing error on each mesh node is then computed as $e\left(\phi(\mathbf{m}_{ij})\right)=|\phi(\mathbf{m}_{ij})-\phi_{\text{exact}}(\mathbf{m}_{ij})|$. To characterize the convergence behavior, we increase the number of particles by decreasing the spacing $h$. For the Wendland and Gaussian kernels, we simultaneously increase the fraction $\epsilon/h$  according to Table \ref{tab:p2msmoothingfactors} as required \cite{raviart1985analysis}. The maximum error over all mesh nodes in a narrow-band of radius $\frac{w}{2}=\frac{1}{16}$ around the zero level-set is plotted in Fig.~\ref{fig:p2mconvergence}.

\begin{table}[ht]
    \centering
    \caption{Smoothing factors $\epsilon/h$ used for the Wendland C2 and Gaussian kernels in the remeshing convergence test.}
    \begin{tabular}{c|c|c|c|c|c|c|c}
         $h$ & 1/32 & 1/64 & 1/128 & 1/256 & 1/512 & 1/1024 & 1/2048 \\\hline
         $\epsilon/h$ & 1.3 & 2.0 & 3.0 & 4.0 & 5.0 & 6.0 & 6.0 \\
    \end{tabular}
    \label{tab:p2msmoothingfactors}
\end{table}

As expected, the $\Lambda_{4,4}$ kernel in the basic formulation (Eq~\eqref{eq:lambda44interpol}) does not converge with its theoretical order of four, as it is derived from moment conservation laws that do not hold for level-set functions. In fact, it diverges slowly, which is 
expected since the maximum error occurs in the most irregular particle neighborhood with an irregularity proportional to the total number of randomly perturbed particles in the domain. Convergence is restored with order 0.81 (still far from the theoretical 4) when renormalizing function values by the amount of contributions they experienced. If in addition to renormalizing function values, the contributions from individual particles are weighted by respective particle volumes, and the smoothing lengths of the kernel functions are increased sufficiently, convergence of order 1.3 is achieved for the Wendland C2 and Gaussian kernels. For these two, however, there is a noticeable effect from the irregular particle distribution, as they are expected to converge with order two for particles distributed on a regular Cartesian grid.

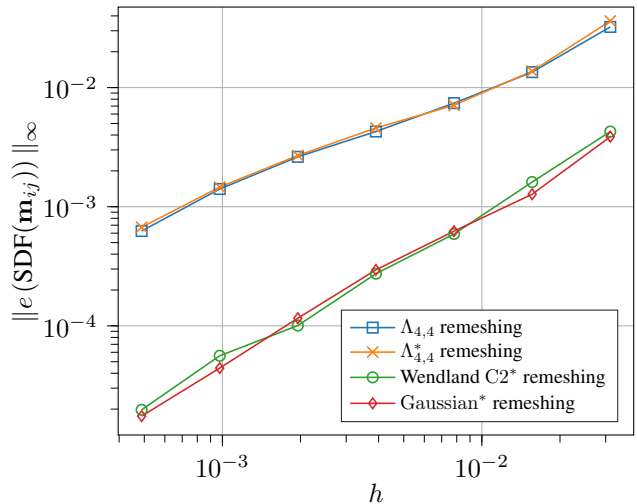
\begin{figure}[ht]
    \centering
    \begin{tikzpicture}

\definecolor{crimson2143940}{RGB}{214,39,40}
\definecolor{darkgray176}{RGB}{176,176,176}
\definecolor{darkorange25512714}{RGB}{255,127,14}
\definecolor{forestgreen4416044}{RGB}{44,160,44}
\definecolor{steelblue31119180}{RGB}{31,119,180}

\begin{axis}[
log basis x={10},
log basis y={10},
tick align=outside,
tick pos=left,
x grid style={darkgray176},
xlabel={$h$},
xmajorgrids,
xmin=0.000396607615408318, xmax=0.0384732629170286,
xmode=log,
xtick style={color=black},
y grid style={darkgray176},
ylabel={$\|e\left(\text{SDF}(\mathbf{m}_{ij})\right)\|_\infty$},
ymajorgrids,
ymin=1.21278391626086e-05, ymax=0.0474500733939735,
ymode=log,
ytick style={color=black},
legend style={legend cell align=left, align=left, draw=white!15!black, nodes={scale=0.75, transform shape}, mark options={scale=1.1}},
legend pos=south east
]
\addplot [semithick, steelblue31119180, mark=square, mark size=2, mark options={solid}]
table {
0.03125 0.03232483824298312824
0.015625 0.01348952520562625955
0.0078125 0.00740729489555021561
0.00390625 0.00427987873804267817
0.001953125 0.00262488891138105192
0.0009765625 0.00140883073813335893
0.00048828125 0.00062541465284596340
};
\addlegendentry{$\Lambda_{4,4}$ remeshing}
\addplot [semithick, darkorange25512714, mark=x, mark size=3, mark options={solid}]
table {
0.03125 0.03622669169230381681
0.015625 0.01375422890649236236
0.0078125 0.00711424822986631532
0.00390625 0.00458064739942115867
0.001953125 0.00269294094371352270
0.0009765625 0.00146221129943255876
0.00048828125 0.00067263805636508158
};
\addlegendentry{$\Lambda_{4,4}^*$ remeshing}
\addplot [semithick, forestgreen4416044, mark=o, mark size=2, mark options={solid}]
table {
0.03125 0.00427663
0.015625 0.00161161
0.0078125 0.000590267
0.00390625 0.000274095
0.001953125 0.000100599
0.0009765625 5.61073e-05
0.00048828125 1.97328e-05
};
\addlegendentry{Wendland $\mathrm{C2}^*$ remeshing}
\addplot [semithick, crimson2143940, mark=diamond, mark size=2, mark options={solid}]
table {
0.03125 0.00386888
0.015625 0.00127353
0.0078125 0.000623578
0.00390625 0.000295154
0.001953125 0.000115922
0.0009765625 4.4212e-05
0.00048828125 1.74745e-05
};
\addlegendentry{$\mathrm{Gaussian}^*$ remeshing}
\end{axis}

\end{tikzpicture}
    \caption{Convergence of the maximum error in the SDF after mesh-based CP redistancing of the remeshed level-set values of the 2D ellipse using different remeshing schemes. The star in the legend entries indicates that the respective remeshing formulation is renormalized.}
    \label{fig:p2mcpconvergence}
\end{figure}

We next test if CP redistancing applied to the interpolated mesh values can restore an overall higher order of convergence. We therefore plot the maximum error in the approximated SDF over the same narrow-band in Fig.~\ref{fig:p2mcpconvergence}. The reference SDF values were obtained using the method outlined in Ref.~\cite{eberly2011distance}.
CP redistancing is done using fourth-order polynomials with a monomial basis of total degree (referred to as ``Taylor 4'' in Ref.~\cite{saye2014}), such that we  expect fifth-order convergence overall~\cite{saye2014}. Due to the intermediate remeshing step, however, the errors converge slower than if redistancing was based on error-free mesh data. When remeshing with the standard or renormalized $\Lambda_{4,4}$ method, the overall convergence is of order 0.95 and 0.96, respectively. When remeshing with the Wendland C2 and Gaussian kernels, overall convergence in the SDF error is of order 1.3. 

In all cases where the remeshing itself converges, the overall redistancing accuracy is therefore limited by the particle-mesh interpolation error. An interesting exception is the $\Lambda_{4,4}$ kernel without renormalization, which does not converge by itself but convergence is restored by CP redistancing. This is because CP redistancing uses a fixed number of grid layers around the surface, rather than a fixed narrow-band width. If the absolute remeshing error is evaluated in a fixed number of grid layers, rather than a fixed-size narrow-band, convergence of order 0.9 is also seen for the remeshing alone. This is because the level-set function has  smaller absolute values closer to the surface. The relative error per mesh node nevertheless remains constant. 

In summary, first interpolating from Lagrangian particles to a regular Cartesian mesh prevents mesh-based CP redistancing from reaching its design order of convergence as particle-mesh interpolation limits the overall accuracy.

\subsubsection{Particle closest-point redistancing}
We compare the above results with the present PCP redistancing method for the same ellipse example with the same irregular particle distribution ($\alpha=0.3$). We use a fourth-order minter basis and a cutoff radius of $r_c=2.5h$.
We set the tolerance $\varepsilon=10^{-14}$ and the maximum number of iterations $k_{\mathrm{max}}=1000$. For the simple ellipse geometry, both the iterative projections and the Newton algorithm converge quickly, i.e.~in 2 to 4 iterations, such that $k_{\mathrm{max}}$ has no effect on the results.

\begin{figure}[ht]
    \centering
    \begin{tikzpicture}[scale=0.97]

\definecolor{darkgray176}{RGB}{176,176,176}
\definecolor{steelblue31119180}{RGB}{31,119,180}

\begin{axis}[
log basis x={10},
log basis y={10},
tick align=outside,
tick pos=left,
x grid style={darkgray176},
xlabel={$h$},
xmajorgrids,
xmin=0.000396607615408318, xmax=0.0384732629170286,
xmode=log,
xtick style={color=black},
y grid style={darkgray176},
ylabel style={yshift=0.2cm},
ylabel={$\|e\left(\text{SDF}\left(\mathbf{x}_p\right)\right)\|_\infty$},
ymajorgrids,
ymin=1.0e-14, ymax=1.0e-06,
ymode=log,
ytick style={color=black},
ytick={1e-15,1e-14,1e-13,1e-12,1e-11,1e-10,1e-09,1e-08,1e-07,1e-06,1e-05,0.0001},
yticklabels={
  \(\displaystyle {10^{-15}}\),
  \(\displaystyle {10^{-14}}\),
  \(\displaystyle {10^{-13}}\),
  \(\displaystyle {10^{-12}}\),
  \(\displaystyle {10^{-11}}\),
  \(\displaystyle {10^{-10}}\),
  \(\displaystyle {10^{-9}}\),
  \(\displaystyle {10^{-8}}\),
  \(\displaystyle {10^{-7}}\),
  \(\displaystyle {10^{-6}}\),
  \(\displaystyle {10^{-5}}\),
  \(\displaystyle {10^{-4}}\)
}
]
\addplot [semithick, black, mark=square, mark size=2, mark options={solid}]
table {
0.03125 5.8979597872266e-07
0.015625 2.567108427221e-08
0.0078125 7.1928845519e-10
0.00390625 3.076531391e-11
0.001953125 3.07509226e-12
0.0009765625 3.32772e-14
0.00048828125 8.488696e-14
};
\end{axis}

\end{tikzpicture}
    \caption{Convergence of the SDF computed by the present particle closest-point (PCP) method for the 2D ellipse case with irregularly ($\alpha=0.3$) distributed particles and fourth-order minter regression polynomials. The observed convergence order is 4.8 (theoretical: 5) with a numerical solver tolerance of $10^{-14}$.}
    \label{fig:pcpellipse}
\end{figure}
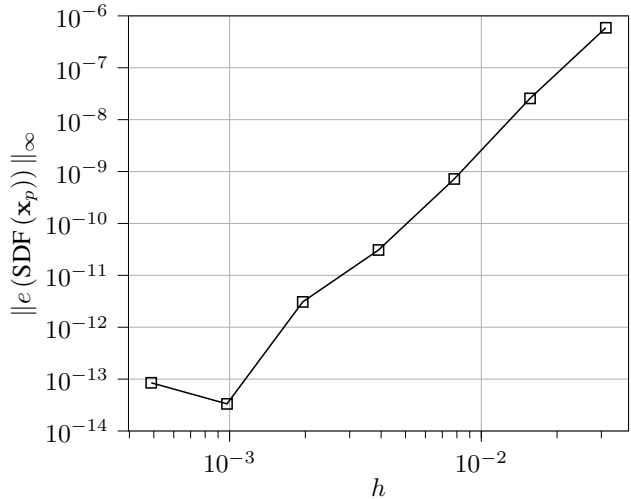

Fig.~\ref{fig:pcpellipse} shows the convergence in the SDF. In absence of any grid, the errors are now evaluated directly on the query particles $P_q$ in the entire narrow-band.
Even for coarse resolutions the error is orders of magnitude smaller than any accuracy achieved by remeshed CP redistancing. The theoretical fifth-order convergence is approximately reached, with a measured order of 4.8, until the error plateaus near the set tolerance $\varepsilon=10^{-14}$.

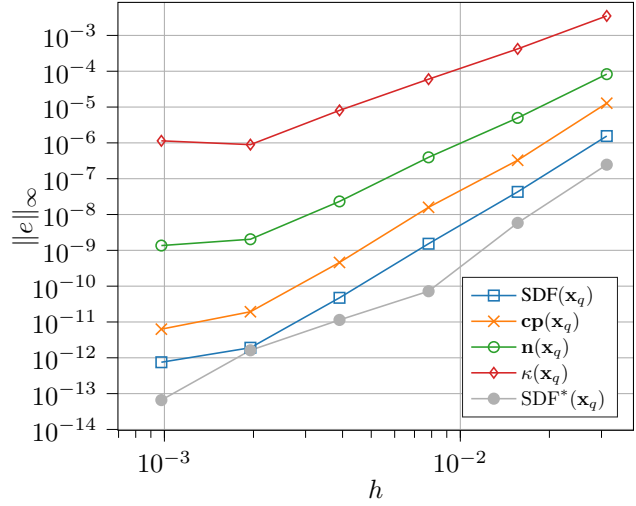
\begin{figure}[ht]
    \centering
    \begin{tikzpicture}

\definecolor{crimson2143940}{RGB}{214,39,40}
\definecolor{darkgray176}{RGB}{176,176,176}
\definecolor{darkorange25512714}{RGB}{255,127,14}
\definecolor{forestgreen4416044}{RGB}{44,160,44}
\definecolor{steelblue31119180}{RGB}{31,119,180}

\begin{axis}[
log basis x={10},
log basis y={10},
tick align=outside,
tick pos=left,
x grid style={darkgray176},
xlabel={$h$},
xmajorgrids,
xmin=0.000696607615408318, xmax=0.0384732629170286,
xmode=log,
xtick style={color=black},
y grid style={darkgray176},
ylabel={$\|e\|_\infty$},
ymajorgrids,
ymin=9.44450491991107e-15, ymax=8.35871871896137e-03,
ymode=log,
ytick style={color=black},
ytick={1e-15,1e-14,1e-13,1e-12,1e-11,1e-10,1e-09,1e-08,1e-07,1e-06,1e-05,0.0001,0.001},
yticklabels={
  \(\displaystyle {10^{-15}}\),
  \(\displaystyle {10^{-14}}\),
  \(\displaystyle {10^{-13}}\),
  \(\displaystyle {10^{-12}}\),
  \(\displaystyle {10^{-11}}\),
  \(\displaystyle {10^{-10}}\),
  \(\displaystyle {10^{-9}}\),
  \(\displaystyle {10^{-8}}\),
  \(\displaystyle {10^{-7}}\),
  \(\displaystyle {10^{-6}}\),
  \(\displaystyle {10^{-5}}\),
  \(\displaystyle {10^{-4}}\),
  \(\displaystyle {10^{-3}}\)
},
legend style={legend cell align=left, align=left, draw=white!15!black, nodes={scale=0.75, transform shape}, mark options={scale=1.1}},
legend pos=south east
]
\addplot [semithick, steelblue31119180, mark=square, mark size=2, mark options={solid}]
table {
0.03125 1.55e-6
0.015625 4.29e-8
0.0078125 1.53e-9
0.00390625 4.73e-11
0.001953125 1.91e-12
0.0009765625 7.52e-13
};
\addlegendentry{SDF$(\mathbf{x}_q)$};

\addplot [semithick, darkorange25512714, mark=x, mark size=3, mark options={solid}]
table {
0.03125 1.28e-5
0.015625 3.27e-7
0.0078125 1.59e-8
0.00390625 4.59e-10
0.001953125 1.93e-11
0.0009765625 6.31e-12
};
\addlegendentry{$\mathbf{cp}(\mathbf{x}_q)$};

\addplot [semithick, forestgreen4416044, mark=o, mark size=2, mark options={solid}]
table {
0.03125 8.29e-5
0.015625 4.98e-6
0.0078125 3.95e-7
0.00390625 2.32e-8
0.001953125 2.04e-9
0.0009765625 1.36e-9
};
\addlegendentry{$\mathbf{n}(\mathbf{x}_q)$};

\addplot [semithick, crimson2143940, mark=diamond, mark size=2, mark options={solid}]
table {
0.03125 3.5e-3
0.015625 4.2e-4
0.0078125 6.0e-5
0.00390625 8.07e-6
0.001953125 8.93e-7
0.0009765625 1.14e-6
};
\addlegendentry{$\kappa(\mathbf{x}_q)$}

\addplot [semithick, darkgray176, mark=*, mark size=2, mark options={solid}]
table {
0.03125 2.46e-7
0.015625 5.84e-9
0.0078125 7.22e-11
0.00390625 1.14e-11
0.001953125 1.62e-12
0.0009765625 6.6e-14
};

\addlegendentry{$\mathrm{SDF}^*(\mathbf{x}_q)$}

\end{axis}

\end{tikzpicture}
    \caption{Convergence of the SDF (observed: 4.9, theoretical: 5), locations of the closest points (observed: 4.8, theoretical: 5), surface normals (observed: 3.8, theoretical: 4), and local mean curvatures (observed: 3.0, theoretical: 3) computed using the PCP method for a 3D ellipsoid discretized on irregular particles ($\alpha=0.3$). $\mathrm{SDF}^*$ is computed using fifth-order polynomials (observed order until $h=1/128$: 5.9, theoretical: 6), whereas all other results use fourth-order polynomials.}
    \label{fig:pcpellipsoid}
\end{figure}

We next assess the convergence of the PCP method for computing derivative surface quantities in a 3D narrow-band around an ellipsoid with semi-major axis $A=0.75$ and both semi-minor axes $B=C=0.5$. We discretize the level-set function on irregularly distributed particles ($\alpha=0.3$) within a narrow-band of width $w=12h$ and use a tolerance of $\varepsilon=10^{-14}$. The maximum error is reported over the entire narrow-band in Fig.~\ref{fig:pcpellipsoid} for fourth- and fifth-order minter regression. For the fourth-order polynomials a cutoff radius of $r_c=2.4h$ is used, for the fifth-order polynomials $r_c=2.6h$.

The theoretical convergence orders are almost achieved when using fourth-order minter regression: the SDF converges with order 4.9, the closest-point transform $\mathbf{cp}(\mathbf{x}_p)$ with 4.8, the surface normals $\mathbf{n}(\mathbf{x}_p)$ with 3.8, and the local curvature $\kappa(\mathbf{x}_p)$ with 3.0 until they plateau at the highest resolution. The SDF and closest-point function converge with the same order and are separated by a constant offset. Since the normal field and local mean curvatures are computed as the derivatives of the SDF, they are third- and second-order polynomials, respectively, with theoretical convergence orders of 4 and 3.

Using fifth-order polynomials increases both accuracy and convergence order (to 5.9) especially for the lower resolutions between $h=\frac{1}{32}$ until $h=\frac{1}{128}$ where round-off errors do not dominate. The convergence order of the other quantities computed with fifth-order PCP are also close to optimal (not shown in the figure to avoid clutter): 6.4 for $\mathbf{cp}(\mathbf{x}_p)$, and 5.2 for $\mathbf{n}(\mathbf{x}_p)$, and 4.0 for $\kappa(\mathbf{x}_p)$.

\begin{figure}[ht]
    \centering
    \begin{tikzpicture}

\definecolor{darkgray176}{RGB}{176,176,176}
\definecolor{darkorange25512714}{RGB}{255,127,14}
\definecolor{steelblue31119180}{RGB}{31,119,180}

\begin{axis}[
log basis y={10},
tick align=outside,
tick pos=left,
x grid style={darkgray176},
xmajorgrids,
xmin=-0.01, xmax=0.505,
xtick style={color=black},
xlabel={$\alpha$},
ylabel={$\|e\left(\text{SDF}\left(\mathbf{x}_p\right)\right)\|_\infty$},
ylabel style={yshift=0.2cm},
y grid style={darkgray176},
ymajorgrids,
ymin=1.0e-11, ymax=1.0e-08,
ymode=log,
ytick style={color=black},
legend style={legend cell align=left, align=left, draw=white!15!black, nodes={scale=0.75, transform shape}, mark options={scale=1.1}},
legend pos=north west
]
\addplot [semithick, steelblue31119180, mark=square, mark size=2, mark options={solid}]
table {
0 2.07e-11
0.01 2.12e-11
0.025 2.33e-11
0.05 3.13e-11
0.1 3.24e-11
0.15 4.3e-11
0.2 4.81e-11
0.25 4.22e-11
0.3 4.73e-11
0.35 4.33e-11
0.4 5.45e-11
0.45 5.66e-11
0.49 6.9e-11
};
\addlegendentry{minter basis}

\addplot [semithick, darkorange25512714, mark=x, mark size=3, mark options={solid}]
table {
0 2.07e-11
0.01 2.12e-11
0.025 2.33e-11
0.05 3.13e-11
0.1 3.24e-11
0.15 4.3e-11
0.2 4.81e-11
0.25 4.22e-11
0.3 4.58e-09
0.35 1.67e-9
0.4 5.58e-09
0.45 2.07e-9
0.49 4.73e-9
};
\addlegendentry{monomial basis}
\end{axis}

\end{tikzpicture}
    \caption{Maximum error of the SDF for the 3D ellipsoid case with $h=\frac{1}{256}$ for increasing irregularity of the particle distribution from a regular grid $\alpha=0$ to the maximum possible irregularity. Results are computed using fourth-order bases for both minter regression and regression using monomial basis functions.}
    \label{fig:canonicalvminter}
\end{figure}
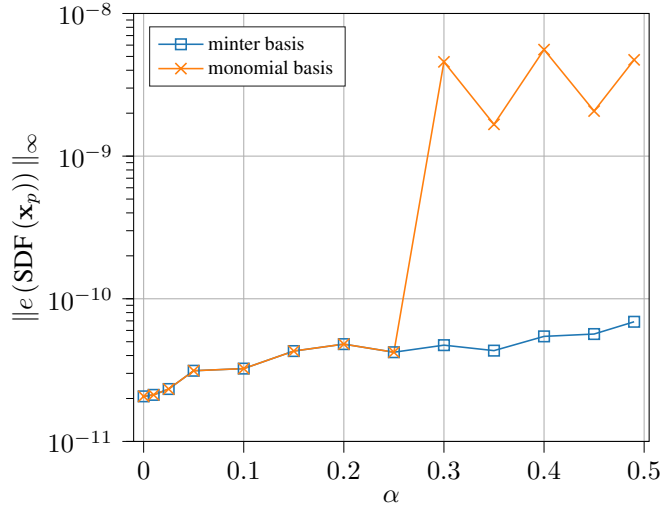

To test the robustness of the PCP method to distortions in the particle distribution, we vary the shift amplitude $\alpha$ and compare the results computed using a minter basis with baseline results obtained using a monomial basis. For both, we use fourth-order polynomials of total degree and a cutoff radius of $r_c=2.4h$. We consider the same ellipsoid as before with resolution fixed to $h=\frac{1}{256}$.
Fig.~\ref{fig:canonicalvminter} shows the maximum error of the SDF within the narrow-band for $\alpha$ from 0 (i.e., regular Cartesian grid) to 0.49. As expected, the errors grow for increasing irregularity of the particle distribution. Comparing the minter basis with the monomial basis, the errors start to differ at $\alpha=0.25$. Beyond this point, error on the monomial basis grows by two orders of magnitude as the regression problem becomes ill-conditioned. This demonstrates that indeed the minter basis and Chebyshev-Lobatto subgrids regularize the regression problem sufficiently to allow convergent redistancing on highly distorted particle distributions.

Finally, we confirm that the PCP method behaves as expected when applied to non-smooth surfaces. For this, we perform PCP redistancing for a rounded rectangle ($C^1$) and a square ($C^0$). On irregular particle distributions ($\alpha=0.3$), the SDF computed using the PCP method converges with order 2 for the rounded rectangle and order 1 for the square. These are the same convergence orders as for the mesh-based CP method \cite{saye2014}, agnostic towards the chosen regression basis and order ($>0$), since a smooth polynomial cannot describe the jump in the curvature (rounded rectangle) or the normals (square). Hence, the leading error term is always limited by the smoothness of the shape itself.

\subsubsection{Parallel scaling}
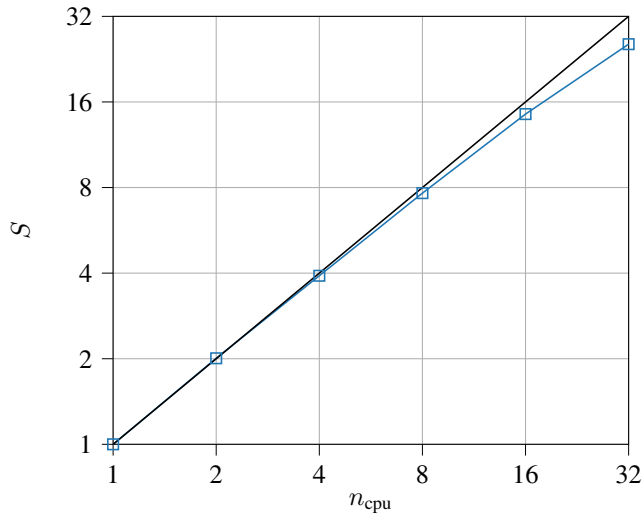
\begin{figure}[ht]
    \centering
    \begin{tikzpicture}

\definecolor{darkgray176}{RGB}{176,176,176}
\definecolor{steelblue31119180}{RGB}{31,119,180}

\begin{axis}[
log basis x={10},
log basis y={10},
xmode=log,
ymode=log,
tick align=outside,
tick pos=left,
x grid style={darkgray176},
xlabel={\(\displaystyle n_{\text{cpu}}\)},
xmajorgrids,
xmin=1.0, xmax=32,
xtick style={color=black},
y grid style={darkgray176},
ylabel={\(\displaystyle S\)},
ymajorgrids,
ymin=1.0, ymax=32,
ytick style={color=black},
ytick={1,2,4,8,16,32},
yticklabels={
  1,
  2,
  4,
  8,
  16,
  32
},
xtick={1,2,4,8,16,32},
xticklabels={
  1,
  2,
  4,
  8,
  16,
  32
}
]
\addplot [semithick, steelblue31119180, mark=square, mark size=2, mark options={solid}]
table {
1 1
2 2.00714285714286
4 3.91637630662021
8 7.64625850340136
16 14.5032258064516
32 25.5454545454545
};
\addplot [semithick, black]
table {
1 1
32 32
};
\end{axis}

\end{tikzpicture}
    \caption{Speedup $S$ over number of processor (CPU) cores $n_{\text{cpu}}$. The solid black line marks the ideal linear speedup.}
    \label{fig:parallelefficiency}
\end{figure}

We test how well our OpenFPM implementation of PCP redistancing scales to multiple processor (CPU) cores. For this, we again use the 3D ellipsoid test case with fixed $h=\frac{1}{512}$, $w=12h$, and $\varepsilon=10^{-14}$ with a fourth-order minter basis and $r_c=2.4h$. We measure the wall-clock time $t_{\text{cpu}}\left(n_{\text{cpu}}\right)$ for different numbers of processor cores $n_{\text{cpu}}$ on an AMD Ryzen Threadripper 3990X CPU with 64 cores and 256 GB of shared memory. We report the parallel speedup 
\begin{equation}
    S(n_{\text{cpu}})=\frac{t_{\text{cpu}}(1)}{t_{\text{cpu}}(n_{\text{cpu}})}
\end{equation}
for this strong scaling (i.e., fixed problem size) in Fig.~\ref{fig:parallelefficiency}, showing near-optimal scalability. For $n_{\text{cpu}}=32$, the parallel efficiency $E(n_{\text{cpu}})=S(n_{\text{cpu}})/n_{\text{cpu}}$ is about 80\%. The 20\% communication overhead are consistent with the volume of the ghost layers required to localize the closest sample particle. 
Another likely reason is an uneven distribution of load amongst the processors.
While we ensured that all processors have the same amount of particles, this does not necessarily result in the same surface area, and therefore $|\mathcal{C}|$, per processor. Hence there is a possible load imbalance during local polynomial regression and surface sampling.

\subsection{Spiraling vortex}
Moving beyond basic geometric shapes, and into dynamically deforming surfaces, we next consider a classic level-set benchmark case in which a circle is stretched and spiraled by an advection velocity field $\mathbf{u}(\mathbf{x})$. The spiraling vortex reaches a state of maximum distortion, and subsequently the velocity field is reversed. At the end, an ideal method would recover the initial circle. In practice, however, numerical errors in the advection of the level-set function and the periodic redistancing accumulate in a final geometry that differs from the initial circle. This difference can be used to compare methods.

For Lagrangian particle methods without remeshing, this test case also introduces another challenge: the particle distribution becomes increasingly distorted and inhomogeneous. The PCP redistancing method is thus confronted with excess information along some directions, and a lack of information along others. This makes an ideal test case to assess the robustness of the PCP method on ill-conditioned particle distributions. 

The initial circle of radius $R=0.15$ centered at $(0.5,0.75)$ in the 2D square domain $[0,1]\times[0,1]$ is discretized with a narrow-band ($w=40h$) of particles. The particles then move with the advection velocity
\begin{equation}
    \mathbf{u}(\mathbf{x},t)=2\cos\left(\tfrac{\pi t}{8}\right)\left(\begin{matrix}
-\sin^2(\pi x)\sin(\pi y)\cos(\pi y)\\
\sin^2(\pi y)\sin(\pi x)\cos(\pi x)
\end{matrix}\right).
\end{equation}
Time integration uses the explicit fourth-order Runge-Kutta scheme with time-step size $\Delta t=\frac{1}{30}$. The velocity field is reversed at $t=4.0$, such that the simulation ends at $t_{\text{end}}=8.0$. At $t=0$, the level-set function values on all particles are initialized to the exact analytical distance to the circle. PCP redistancing with fourth-order minter regression is done after each time step of the advection with $\varepsilon=10^{-10}$, $r_c=15h$, and $k_{\mathrm{max}}=100$. The reduced maximum number of iterations is required in this case, as the solver is unable to reach the desired tolerance for some of the most distorted particle distributions.

\begin{figure}[ht]
    \centering
    \includegraphics[scale=0.34]{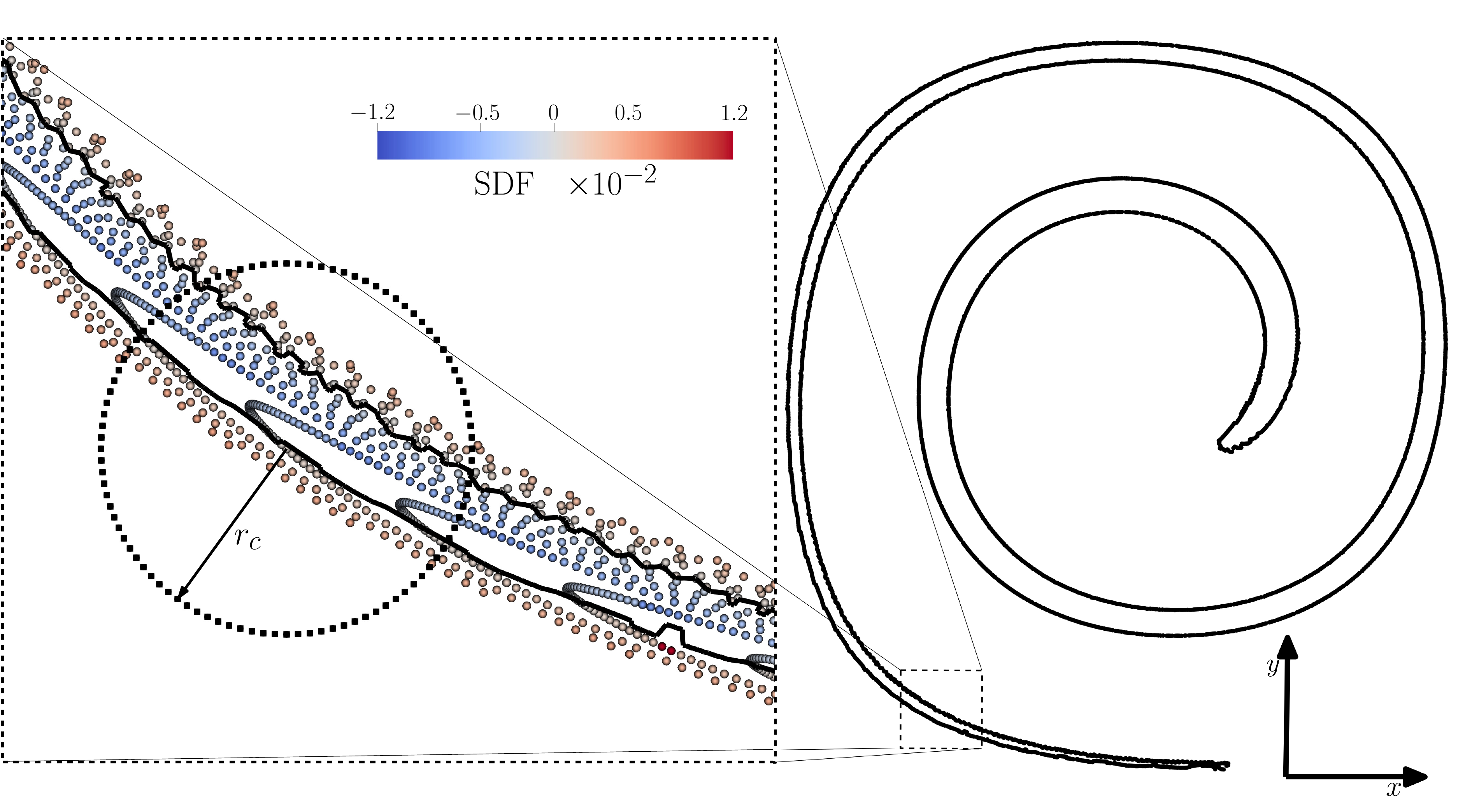}
    \caption{The most ill-conditioned particle distribution in the spiraling vortex case with $h=\frac{1}{512}$ at $t=4.0$. Spheres represent particle positions and are colored level-set function values. The dotted circle is an exemplary regression neighborhood of a particle with $r_c=15h$. The solid line represents the reconstructed surface $\Gamma_t$.}
    \label{fig:vortexdistribution}
\end{figure}

The large cutoff radius reduces the accuracy in the early and late time steps of this example, yet it is beneficial during steps involving extremely ill-conditioned particle distributions, where the particle density along one direction differs significantly from the density in the other direction. The flow field causing particles to align poses severe limitations to polynomial regression approaches and also causes the approximated zero level-set to lose its smoothness during the simulation. The most ill-conditioned particle distribution (at $t=4.0$) is visualized in Fig.~\ref{fig:vortexdistribution}. The zero level-set of the SDF is shown as an isocontour as determined by Paraview \cite{paraview}.

As can be seen in Fig.~\ref{fig:vortexdistribution}, the particle distribution is particularly sparse in the direction orthogonal to the surface. This is challenging for any redistancing scheme, since the SDF solely depends on information in the orthogonal direction. Another challenging aspect of this particle distribution is that multiple zero level-sets are present in each $r_c$-neighborhood, as shown exemplary for one particle in the closeup in Fig.~\ref{fig:vortexdistribution}, dotted circle.

\begin{figure}[ht]
    \centering
    \includegraphics{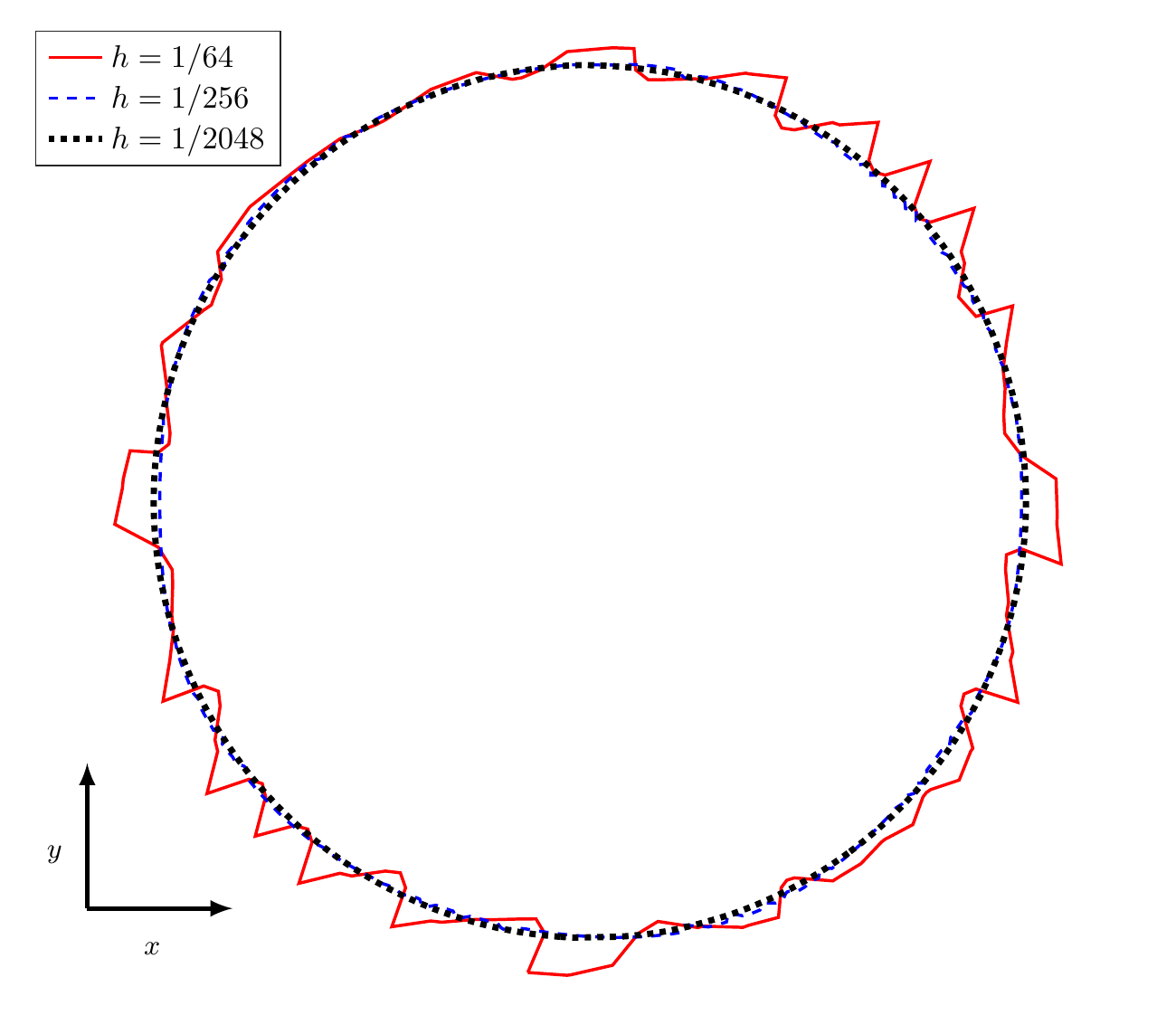}
    \caption{Visualization of the final zero level-set at $t_{\text{end}}=8.0$ for three different inter-particle spacings (line styles, see inset legend).}
    \label{fig:final_circle}
\end{figure}

Running the simulation until the final time $t_{\text{end}}=8.0$, we can observe how well the PCP method with redistancing at each time step is able to recover the initial circle. Fig.~\ref{fig:final_circle} visualizes the final isocontours reconstructed by Paraview \cite{paraview} for three different resolutions. As expected, simulations with fewer particles suffer more. For $h=\frac{1}{64}$, the final level-set resembles a circle on average, but has noticeable errors. These errors do, however, converge with increasing numbers of particles, and the final shape for $h=\frac{1}{2048}$ is visually indistinguishable from the original circle. This can also be seen in the convergence of the average and maximum errors in the SDF and the enclosed area $A$ ($e(A)=|(A-A_{\text{exact}})|/A_{\text{exact}}$) of the level-set as reported in Table \ref{tab:area_convergence}. The average error in the SDF converges with order 1.9, the maximum error converges with order 1.5, and the area converges with order 1.8. 

This confirms that the PCP method converges (albeit with sub-optimal order) despite the severely ill-posed particle distributions occurring during the spiraling vortex dynamics.

\begin{table}[ht]
    \centering
    \caption{Errors in the SDF values and the enclosed area $A$ for the final state of the spiraling vortex.}
    \begin{tabular}{c|c|c|c}
         $h$ & $e(A)$ & $\|e(\text{SDF})\|_2$ & $\|e(\text{SDF})\|_\infty$\\\hline
          1/32 &\num{5.85e-2}&\num{3.60e-2}&\num{8.8e-2}\\
          1/64&\num{3.83e-2}& \num{1.42e-2}&\num{2.4e-2}\\
          1/128&\num{1.18e-2}&\num{2.90e-3}&\num{8.1e-3}\\
          1/256&\num{1.11e-3}&\num{1.10e-3}&\num{2.2e-3}\\
          1/512&\num{3.24e-4}&\num{3.98e-4}&\num{1.1e-3}\\
          1/1024&\num{2.99e-4}&\num{6.90e-5}&\num{4.4e-4}\\
          1/2048&\num{4.04e-5}&\num{1.62e-5}&\num{1.8e-4}\\
    \end{tabular}
    \label{tab:area_convergence}
\end{table}

\subsection{Multi-phase hydrodynamics}\label{sec:resultsSPHmultiphase}
As a final test case, we consider the more ``real-world'' example of droplet dynamics in multi-phase hydrodynamics in both 2D and 3D. In this test case, a closed
surface is coupled to two surrounding fluid phases, one inside and one outside. The curved interface between the two fluids has a surface tension and thus exerts forces on the fluids that depend on its curvature and cause dynamic flows, which in turn again advect and shape the interface. Both fluids are modeled using the incompressible Navier-Stokes equations
\begin{gather}
    \frac{\text{D}\rho}{\text{D}t}=-\rho\nabla\cdot\mathbf{u}\, ,\label{eq:contiequation}\\
    \frac{\text{D}\mathbf{u}}{\text{D}t}=-\frac{1}{\rho}\nabla p+\frac{1}{\rho}\eta\Delta\mathbf{u} + \mathbf{F}^{(s)}\, ,\label{eq:momentumeq}
\end{gather}
where density is denoted by $\rho$, pressure by $p$, $\eta$ is the dynamic viscosity, and $\Delta$ is the Laplace operator. The surface tension effect is modeled through a continuum surface force \cite{brackbill1992continuum}. Here, a volumetric surface force is active in the smoothed out interface, or transition region, which can be described with a smooth surface delta function $\delta$. The surface force then reads
\begin{equation}\label{eq:csfmodel}
\mathbf{F}^{(s)}=-\frac{\tau}{\rho}\kappa\mathbf{n}\delta
\end{equation}
and acts in the normal direction of the surface and is proportional to both the surface tension $\tau$ and the local mean curvature $\kappa$.

The continuity Eq.~\eqref{eq:contiequation} and momentum Eq.~\eqref{eq:momentumeq} are complemented by the Cole equation of state \cite{cole1948underwater}
\begin{equation}\label{eq:eos}
    p(\rho)=\frac{c^2\rho_0}{\gamma}\left(\left(\frac{\rho}{\rho_0}\right)^{\!\!\gamma}-1\right),
\end{equation}
linking the pressure $p$ to the density $\rho$ via a reference density $\rho_0$, a speed of sound $c$ and the polytropic index of the fluid $\gamma$. We choose the speed of sound $c=100$ at least one order of magnitude larger than the maximum flow velocity $\|\mathbf{u}_\mathrm{max}\|_2$ to ensure Mach numbers $Ma:=\|\mathbf{u}_\mathrm{max}\|_2/c\lesssim 0.3$ remain in the region of incompressible flow. We further set $\gamma=7$ and consider two identical fluids with $\rho_0=1.0$ and $\eta=0.5$, separated by an interface with $\tau=50.0$.

We solve the Navier-Stokes equations inside and outside the droplet without the interfacial forces $\mathbf{F}^{(s)}$ using a weakly compressible Smoothed Particle Hydrodynamics (SPH) approach \cite{adami2010new}. Further details of the numerical method and the test case are given in Appendix~\ref{sec:sphmultiphaseflow}.
The interfacial forces $\mathbf{F}^{(s)}$ in Eq.~\eqref{eq:csfmodel} are computed in two ways for comparison: (1) using the present PCP method to compute the SDF, the surface normals, and the mean curvatures, resulting in Eq.~\eqref{eq:pcpsurfaceforce}, and (2) using a colorfield function and SPH operators \cite{morris2000simulating}, resulting in Eq.~\eqref{eq:sphsurfaceforce}. For the quantitative analysis we first consider a 2D model. Here, we perform simulations for three different resolutions, $h=\frac{1}{32}$, $\frac{1}{64}$, and $\frac{1}{128}$ with a smoothing factor of $\epsilon/h=3$. The time steps for the three different resolutions are $\Delta t=\num{2e-4}$, $\Delta t=\num{1e-4}$, and $\Delta t=\num{5e-5}$.

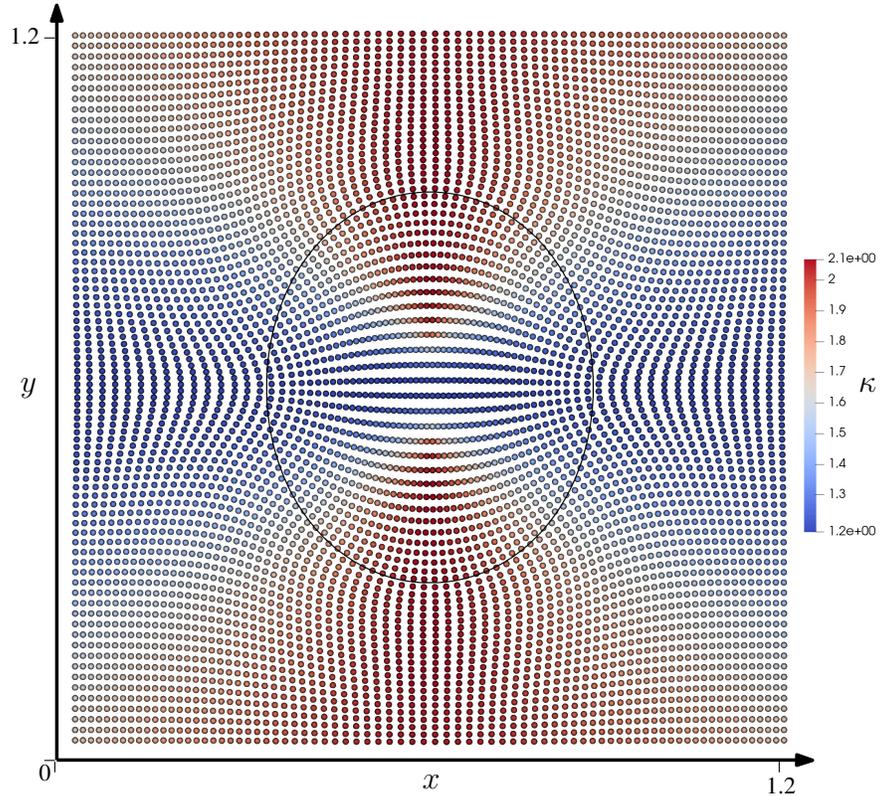
\begin{figure}[ht]
\centering
\begin{picture}(300,300)
    \centering
    \put(0,0){\includegraphics[scale=0.26]{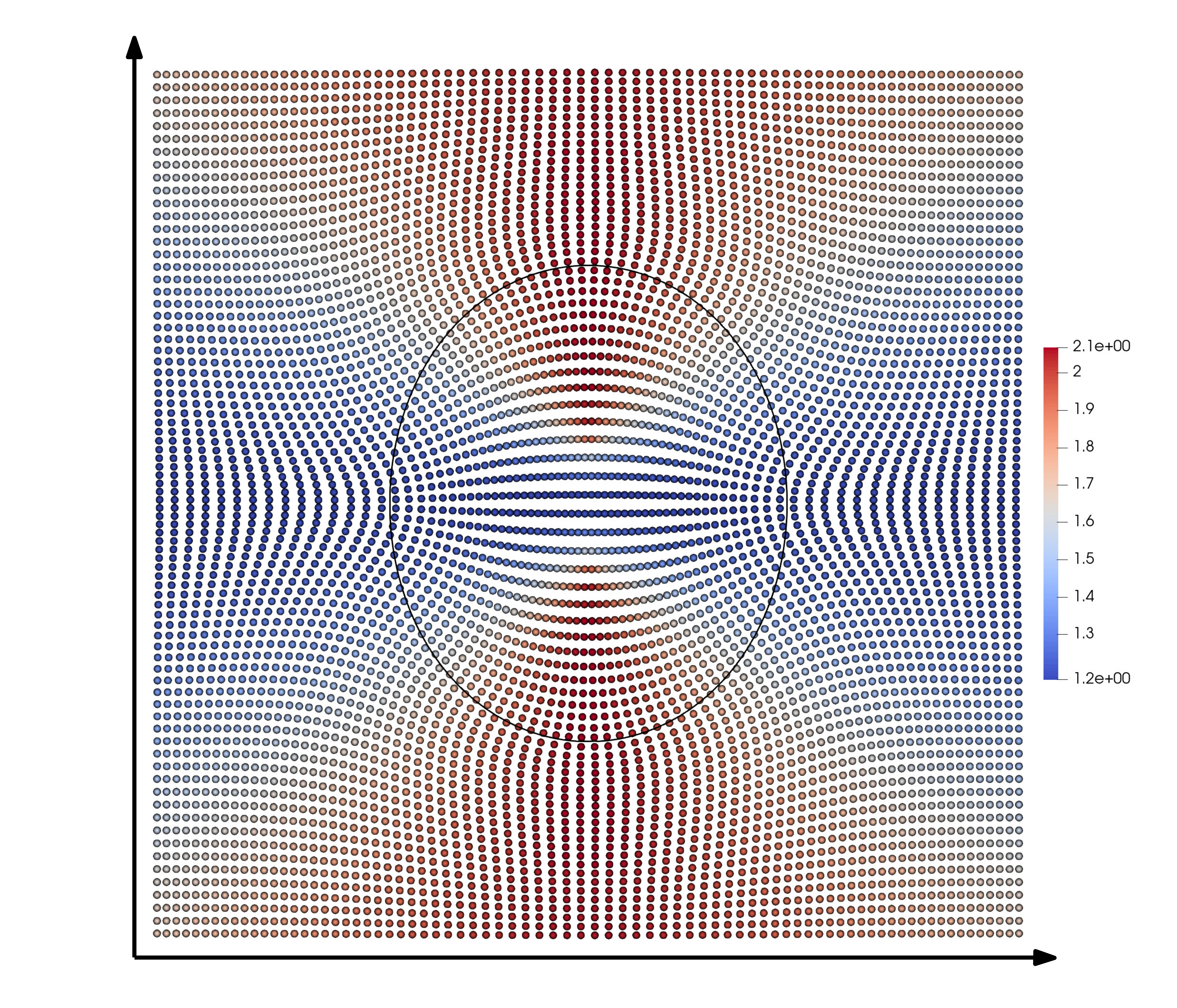}}
    \put(180,5){\large $x$}
    \put(28,155){\large $y$}
    \put(35,8){\footnotesize 0}
    \put(310,3){\footnotesize 1.2}
    \put(315,11){\line(0,3){4}}
    \put(37,289){\line(1,0){4}}
    \put(37,15.5){\line(1,0){4}}
    \put(41,11){\line(0,1){4}}
    \put(24,287){\footnotesize{1.2}}
    \put(345,155){\large $\kappa$}
\end{picture}
\caption{Computed mean curvature values of the closest points (color bar) at $t\approx0.15$ for $h=\frac{1}{32}$ for the two-phase droplet hydrodynamics. The zero level-set in solid black was reconstructed by Paraview \cite{paraview} from the SDF values computed using the PCP method.}
\label{fig:curvature_distr_multiphase}
\end{figure}

Initially, at time $t=0$, both fluid phases are at rest (zero velocity everywhere) and the interface is an ellipse (Eq.~\eqref{eq:ellipseeq}) with $A=0.75$ and $B=0.5$ embedded in the 2D domain $\left(-1.22,1.22\right)\times\left(-1.22,1.22\right)$ with periodic boundary conditions in both directions. Interface regions of higher curvature exert greater force than those of lower curvature, causing oscillating deformation of the incompressible droplet, eventually subsiding with the shape of minimal surface area, a circle. For the 3D case, this is visualized in the Supplementary Video of the web version of this article. We run the simulations until $t_{\text{end}}=1$. Fig.~\ref{fig:curvature_distr_multiphase} visualizes the particle distribution, the computed mean curvature values, and the zero level-set reconstructed from the advected level-set values at $t\approx0.15$, when the droplet assumes its maximum vertical elongation.

\begin{figure}[ht]
\centering
\begin{picture}(300,300)
    \centering
    \put(35,9){\includegraphics[scale=0.26]{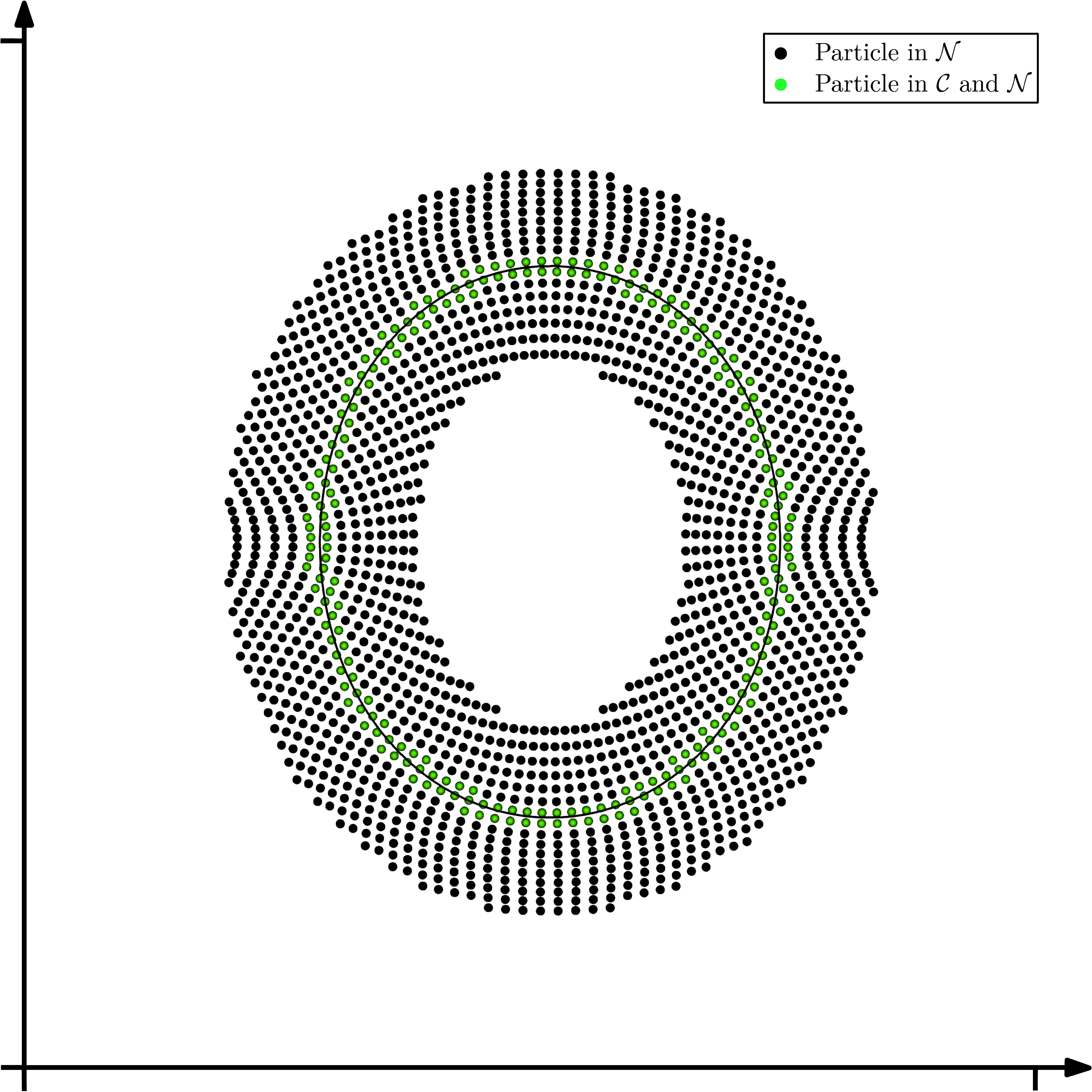}}
    \put(180,5){\large $x$}
    \put(28,155){\large $y$}
    \put(35,8){\footnotesize 0}
    \put(307.5,2.5){\footnotesize 1.2}
    \put(23,288){\footnotesize{1.2}}
\end{picture}
\caption{Narrow-band of width $w=15h$ at $t\approx0.15$ for the two-phase droplet hydrodynamics test case with $h=\frac{1}{32}$. Close particles used as seeds to generate sample particles during redistancing are shown in green. The solid black line shows the interface surface reconstructed by Paraview \cite{paraview} based on the SDF values computed using the PCP method.}
\label{fig:narrowband_surfflag_multiphase}
\end{figure}

Table~\ref{tab:wallclocktime} compares the computational times of the entire 2D simulation using the present PCP method and using colorfield-SPH for different resolutions $h$ with the same $\Delta t=\num{5e-5}$. 
Computing geometric quantities involves more computations per particle in the PCP method than in the colorfield-SPH approach, which is why the computational time is not smaller for PCP at low spatial resolution, even though the computations are confined to the narrow band. However, as the fraction of particles that are inside the narrow band decreases with decreasing average inter-particle spacing $h$, the computational cost of PCP reduces for finer resolutions compared to SPH. Fig.~\ref{fig:narrowband_surfflag_multiphase} shows the narrow band of width $w=15h$, to which geometric computations are confined, for $h=1/32$.

\begin{table}[ht]
    \centering
    \caption{Wall-clock times in minutes required for the entire 2D two-phase flow simulation with different spatial resolutions $h$ using a fixed time resolution of $\Delta t=\num{5e-5}$ and two different geometric computing approaches for the interface dynamics: the colorfield-SPH approach (SPH) and the present particle closest-point (PCP) method. All simulations are performed on a single processor core of an AMD Ryzen Threadripper 3990X.}
    \begin{tabular}{c|c|c|c}
         $h$ & 1/32 & 1/64 & 1/128 \\\hline
         SPH & 71 & 272 & 1061 \\
         PCP & 72 & 201 & 646 \\
    \end{tabular}
    \label{tab:wallclocktime}
\end{table}

A key advantage of a closest-point approach can be seen in Figs. \ref{fig:curvature_distr_multiphase} and \ref{fig:stdkappa} (\ref{sec:appfigures}): geometric quantities are computed at the closest point, i.e. on the surface itself, and then extended along the surface normal to particles in the embedding space. For finite $h$, this differs from the the colorfield-SPH approach, which computes geometric quantities of the isocontours of the level-set function on which the particles lie. Hence, it is unsurprising that in Fig.~\ref{fig:stdkappa} in \ref{sec:appfigures} the standard deviation of the computed mean curvature values is smaller for the PCP method. The minimum standard deviation of the mean curvature values occurs when the interface becomes a circle. For the colorfield-SPH approach in the highest resolution, the smallest achieved standard deviation is 0.105, 
while the values range from \num{8e-4} to \num{3e-4} for the different resolutions in the PCP approach. This is orders of magnitude better than the colorfield approach and generally satisfying, given that the results are limited by the time resolution of the simulations, and the exact moment the geometry becomes circular may not be captured.
A further limiting factor of the results can be identified in the accuracy of the SPH approximation of the remaining terms in the Navier-Stokes equations for the surrounding fluids.

Looking at the average mean curvature across particles in the interface region (Fig.~\ref{fig:meankappa}, \ref{sec:appfigures}) we notice that the PCP method is slightly biased towards higher curvatures than the exact value an ideally incompressible fluid would yield. The volume reconstructed from the computed SDF is consistent with the mean curvature. For $h=\frac{1}{128}$, the volume reconstructed from the SDF computed by PCP has a relative error of -0.5\% at the end of the simulation after a total of 40,000 redistancing steps (computing the volume using colorfield SPH is not straightforward). Given that the results from the colorfield-SPH approach have a similar bias for lower resolutions, but are less accurate, the bias is probably due to the simulated fluid not being ideally incompressible in the weakly compressible SPH approach. In summary, PCP not only computes curvature values at the correct iso-contour, but the values are also less noisy than those from colorfield-SPH. This is because it does not use a binary indicator function which heavily depends on the particle distribution in capturing the geometry.

\begin{figure}[ht]
    \centering
    \includegraphics[scale=0.15]{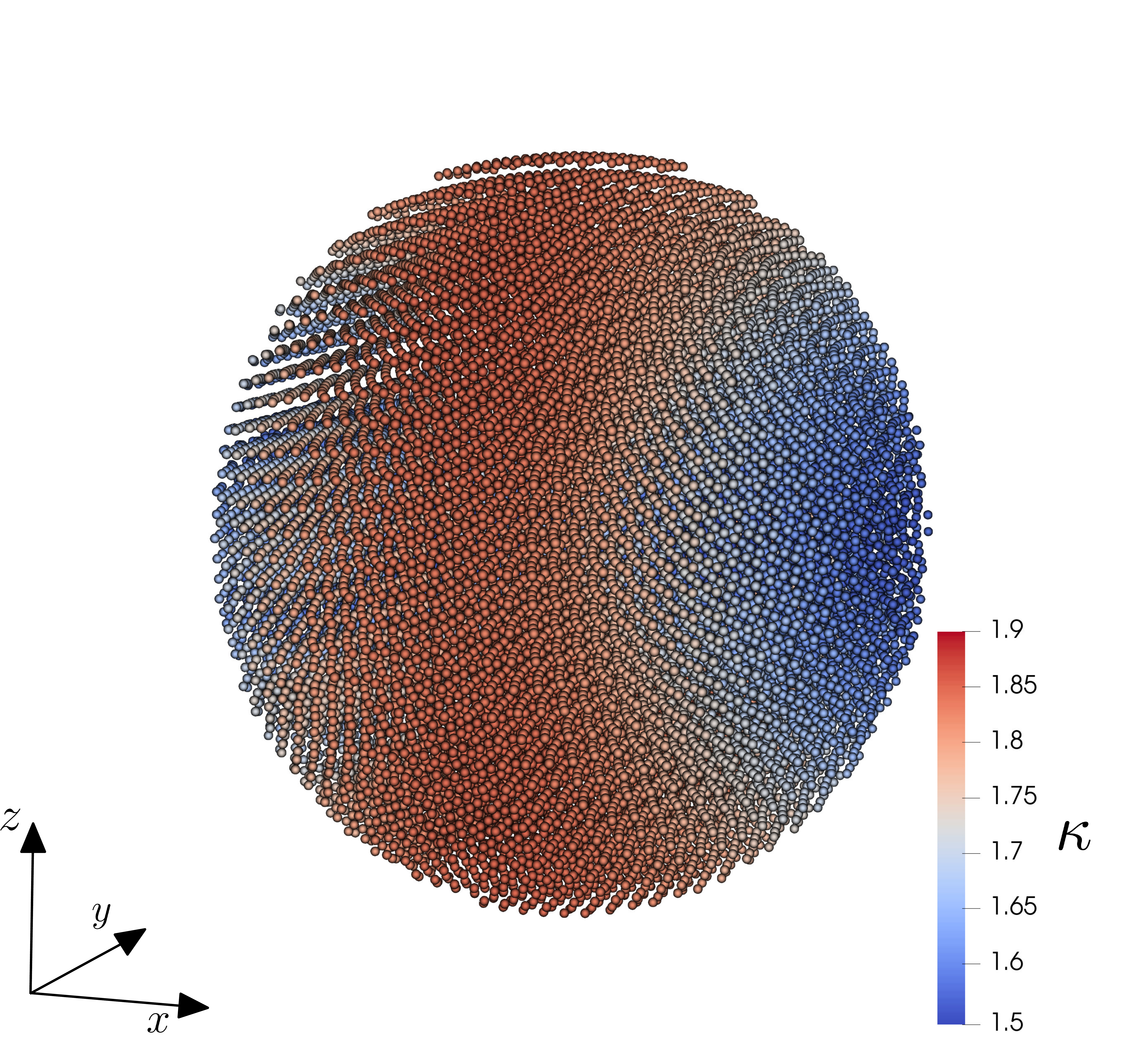}
    \caption{Particles of the inner fluid phase of a 3D oscillating droplet simulation color-coded according to the computed mean curvature values.}
    \label{fig:3Ddroplet}
\end{figure}

Finally, we also test the PCP multi-phase SPH fluid simulation for a 3D oscillating droplet with an initial shape of an ellipsoid with semi-major axis $A=0.75$ and both semi-minor axes $B=C=0.5$. We use the same parameters as in the 2D case with $h=1/32$. To lessen the computational burden stemming from the SPH operators, we reduce the smoothing factor to $\epsilon/h=2$, and to alleviate the effort from geometric computing, we reduce the narrow-band width to $w=10H$. Fig. \ref{fig:3Ddroplet} visualizes the particles of the inner fluid phase with their computed mean curvature values during the maximum vertical elongation of the droplet at $t\approx0.19$. We also compute results for a lower viscosity value of $\eta=0.3$, visualized in the Supplementary Video of the web version of this article.

Similarly as in the 2D simulations, the PCP method computes smooth mean curvature values of the dynamically deforming surface in 3D. Also in 3D, the PCP results are more accurate and computationally less expensive than the colorfield-SPH approach. This suggests that the PCP method is well suited for simulations involving dynamically deforming shapes.

\section{Discussion and Outlook}
We presented a higher-order redistancing scheme for fully Lagrangian particle level-set methods, extending closest-point redistancing \cite{saye2014} to irregularly distributed points. Unlike previous particle level-set methods \cite{hieber2005lagrangian,enright2002hybrid}, the proposed approach does not require any form of interpolation from particles to mesh nodes, which we have shown to limit convergence for non-conserved level-set functions. The presented Particle Closest-Point (PCP) method relies on minter regression on Chebyshev-Lobatto subgrids to achieve numerical robustness. We have shown that this renders the method more robust to distortion in the particle distribution than regression using monomial bases.

In the PCP method, the particles act as sample points of the level-set function, which is in contrast to colorfield approaches where particles directly represent the (smoothed) presence of a certain phase~\cite{morris2000simulating,adami2010new,marrone2010fast}. Our approach is purely geometric and allows for arbitrarily placed query points to be redistanced. Hence, the approach is readily applicable to the initialization of geometric quantities of new particles, be it in multi-resolution methods \cite{reboux2012self} or as a part of remeshing.

We showed that the PCP method provides high-order convergent geometric quantities for basic geometries without requiring any connected mesh. We tested the robustness of the approach by studying highly irregular particle distributions and found that the high-order polynomial regression with a Lagrange basis on unisolvent nodes outperforms monomial regression approaches for irregular particle distributions, and thus is well-suited for a Lagrangian particle level-set method. This was also reflected during the application of the PCP method to ill-conditioned particle distributions in the spiraling vortex case, in which convergence was maintained. Finally, a more complex problem was studied in which an oscillating droplet was simulated in conjunction with SPH multi-phase flow operators. 

Overall, we found that the PCP method copes well with irregular particle distributions, yet we still expect it to require a certain homogeneity in the distribution as it benefited from larger smoothing lengths in the SPH operators. We also observed that some particles approached the zero contour and remained there. To avoid volume loss, we prevented particles from taking on a true 0 as a level-set value and from changing the sign of their level-set values~\cite{enright2002use}. This clearly is a limitation of the proposed approach and some form of particle distribution regularization will eventually be necessary. Another limitation is the non-uniform load distribution when implementing PCP methods on parallel computers. This is because the computational cost on a given processor not only depends on the number of particles it handles (which can be evenly distributed), but also on the local geometry of the surface (which is impossible to predict in a dynamic simulation), through the local regression problems. Nevertheless, the narrow-band character of the PCP approach confines the computational effort of the geometric computing framework to the proximity of the surface. We found that this leads to an overall faster simulation time than a naive colorfield-SPH approach in highly resolved multi-phase flows.

As with most level-set methods, the proposed PCP approach is suitable for continuous, orientable, closed surfaces that do not self-intersect. This is the case for any closed surface that possesses a tubular neighborhood. For non-smooth surfaces, the convergence order of the method is bounded by the smoothness class of the surface and the degree of the regression polynomials used, whichever is smaller. If the smoothness class of the surface is smaller than the used polynomial degree, the leading error term is then determined by the derivative order in which discontinuities first appear.

Future work could further optimize the numerical robustness of the method by testing different approaches to choosing particles that lie close to the unisolvent nodes of a Chebyshev-Lobatto subgrid. Further research could consider how Lagrangian formulations could incorporate interface velocity extensions~\cite{henneaux2021high} to reduce the required frequency of redistancing steps. 

Future applications of the PCP method can take advantage of the polynomial minter regression that is the defining feature of the method. This can, for example, include numerically solving PDEs on surfaces, where constant orthogonal extension is a popular solution approach~\cite{ruuth2008simple,marz2012calculus}. 
There, the PCP method solves two of the main challenges: It accurately computes the CP transform using an orthogonal decomposition of the regression problem, which can straightforwardly be reused to also approximate the values of other fields at the computed closest point. Due to its computational efficiency, parallel scalability, and robustness against distortion in the particle distribution, the PCP method could therefore by key to solving PDEs on dynamically deforming surfaces.

The C++ implementation of the presented particle closest point method in OpenFPM is publicly available from \texttt{https://git.mpi-cbg.de/mosaic/software/parallel-computing/openfpm/openfpm\textunderscore numerics}.

\section*{Acknowledgements}
We thank Dr.~Michael Hecht (Center for Advanced Systems Understanding, G\"orlitz), Alejandra Foggia, Johannes Pahlke, and Justina Stark (all Sbalzarini group) for discussions and proofreading, and Dr.~Pietro Incardona (University of Bonn) for his helpful support in the OpenFPM implementation of the presented method. 
This work was supported by the German Federal Ministry of Education and Research (Bundesministerium f\"{u}r Bildung und Forschung, BMBF) in
the joint project ``6G-life'' (ID 16KISK001K).

\section*{Declaration of competing interest}
The authors have no conflict of interest to declare.


\appendix

\section{Particle-mesh interpolation}\label{sec:appendixkernels}
A popular kernel for particle-mesh interpolation is the $\Lambda_{4,4}$ kernel \cite{cottet2014high}, which conserves the first 4 moments and produces $C^4$ regular results:
\begin{flalign}
\Lambda_{4,4}(q)=
\begin{cases}
1-\frac{5}{4}q^2+\frac{1}{4}q^4-\frac{100}{3}q^5+\frac{455}{4}q^6-\frac{295}{2}q^7+\frac{345}{4}q^8-\frac{115}{6}q^9,&\text{if }0\leq q<1, \\
-199+\frac{5485}{4}q-\frac{32975}{8}q^2+\frac{28425}{4}q^3-\frac{61953}{8}q^4+\frac{33175}{6}q^5 \\
-\frac{20685}{8}q^6+\frac{3055}{4}q^7-\frac{1035}{8}q^8+\frac{115}{12}q^9,&\text{if }1\leq q<2,\\
5913-\frac{89235}{4}q+\frac{297585}{8}q^2-\frac{143895}{4}q^3+\frac{177871}{8}q^4-\frac{54641}{6}q^5 \\
+\frac{19775}{8}q^6-\frac{1715}{4}q^7+\frac{345}{8}q^8-\frac{23}{12}q^9,&\text{if }2\leq q<3, \\
0,&\text{else}.
\end{cases}
\end{flalign}
It can be used to evaluate mesh node values as 
\begin{equation}\label{eq:lambda44interpol}
    \phi_{ij} = \sum_p \phi_p\Lambda_{4,4}\left(\frac{|x_i-x_p|}{h}\right)\Lambda_{4,4}\left(\frac{|y_j-y_p|}{h}\right),
\end{equation}
in which $i$ and $j$ are the grid indices for $x$ and $y$ directions, respectively, and $p$ is the index for the particles.

This classic approach to particle-mesh interpolation is designed for intensive fields of conserved quantities, such as density or concentration fields  \cite{bergdorf2010lagrangian,bergdorf2007multiresolution,hieber2004remeshed}. As level-set functions generally do not obey any conservation laws, the different amounts of contributions individual mesh nodes receive need to be accounted for as soon as the particle distribution becomes irregular. 
This is done by renormalizing the interpolated quantities according to
\begin{align}\label{eq:lambda44renormalizedinterpol}
    \phi_{ij}=
    \left(\sum_p\Lambda_{4,4}\left(\frac{|x_i-x_p|}{h}\right)\Lambda_{4,4}\left(\frac{|y_j-y_p|}{h}\right)\right)^{\!-1}\sum_p \phi_p\Lambda_{4,4}\left(\frac{|x_i-x_p|}{h}\right)\Lambda_{4,4}\left(\frac{|y_j-y_p|}{h}\right).
\end{align}
Similarly, the particle contributions can be scaled by individual volumes to yield the classic particle representation of an arbitrary field as
\begin{equation}\label{eq:particlerep}
    f\left(\mathbf{x}\right)\approx\sum_pf_pW_\epsilon\left(\|\mathbf{x} - \mathbf{x}_p\|_2\right)V_p\, ,
\end{equation}
where $W$ is a local, symmetric, normalized kernel function with a smoothing length of $\epsilon$. The smoothing length determines how many particles contribute to a field evaluation, and it strongly influences the convergence properties of particle methods. Generally, as the inter-particle spacing $h$ becomes smaller, the smoothing length $\epsilon$ should also become smaller, yet the ratio $\frac{\epsilon}{h}$ should grow to ensure convergence of the scheme \cite{raviart1985analysis}.

In Eq.~\eqref{eq:particlerep}, $V_p$ is the volume associated with particle $p$, which can be computed as
\begin{equation}\label{eq:particlevolume}
    V_q=\left(\sum_pW_\epsilon\left(\|\mathbf{x}_q - \mathbf{x}_p\|_2\right)\right)^{\!-1}.
\end{equation}
The particle function approximation in Eq.~\eqref{eq:particlerep} of a level-set function can then also be renormalized and subsequently evaluated at mesh nodes $ij$:
\begin{equation}\label{eq:particlereprenormalized}
    \phi_{ij}=\left(\sum_pW\left(\|\mathbf{m}_{ij} - \mathbf{x}_p\|_2,\epsilon\right)V_p\right)^{\!-1}\sum_p\phi_pW\left(\|\mathbf{m}_{ij} - \mathbf{x}_p\|_2,\epsilon\right)V_p\, .
\end{equation}
In the main text, we compare this classic particle-mesh interpolation scheme with two different kernel functions popular in the SPH community. The first is the Wendland C2 kernel \cite{wendland1995piecewise}
\begin{equation}\label{eq:wendlandc2}
    W_\epsilon\left(q\right)=
    \begin{cases}
    \sigma_{2D}\left(1-\frac{q}{2}\right)^4\left(1+2q\right)\qquad&\text{if } 0\leq q<2,\\
    0&\text{else,}
    \end{cases}
\end{equation}
where the variable $q$ is defined as: $q=\frac{\|\mathbf{x}-\mathbf{x}_p\|_2}{\epsilon}$, and $\sigma_{2D}=\frac{7}{4\pi\epsilon^2}$ is a normalization factor ensuring that the kernel integrates to one.
The second popular SPH kernel is the Gaussian
\begin{equation}
    W_\epsilon\left(q\right)=
    \begin{cases}
    \sigma_{2D}e^{-q^2}\qquad&\text{if }0\leq q\leq3,\\
    0&\text{else,}
    \end{cases}
\end{equation}
with $q$ defined as for the Wendland kernel, and $\sigma_{2D}=\frac{1}{\pi\epsilon^2}$  the normalization constant.

\section{SPH formulation for multi-phase flow}\label{sec:sphmultiphaseflow}

We discretize both continuum fluids with a set of in total $n_p$ particles, 
initialized on a regular Cartesian grid of spacing $h$ covering the entirety of the simulation domain. The masses $m_i=M/n_p$ of the particles are computed by considering the total mass of the fluids $M$, resulting from the reference density and the occupied volume. To estimate the density of a single particle $i$, density summation is performed as
\begin{equation}\label{eq:densitysummation}
    \rho_i=m_i\sum_jW_{ij}\, ,
\end{equation}
where the Wendland C2 kernel from Eq.~\eqref{eq:wendlandc2} is used as $W_{ij}=W_\epsilon(\|\mathbf{x}_i-\mathbf{x}_j\|_2)$. Simultaneously, the volume of each particle $V_i$ is computed according to Eq.~\eqref{eq:particlevolume}. Having computed the densities, the pressures are obtained by evaluating Eq.~\eqref{eq:eos} for $\rho_i$. At $t_0=0$, all velocities are set to $\mathbf{u}_i=\mathbf{0}$. The change in velocity per particle is determined by the discrete momentum equation as \cite{adami2010new}:
\begin{align}
    \frac{\text{D}\mathbf{u}_i}{\text{D}t}=&-\frac{1}{m_i}\sum_j\left(V_i^2+V_j^2\right)\frac{\rho_iP_j+\rho_jP_i}{\rho_i+\rho_j}\nabla W_{ij}+\frac{1}{m_i}\sum_j\eta\left(V_i^2+V_j^2\right)\frac{\mathbf{u}_{ij}}{r_{ij}}\frac{\partial W}{\partial r_{ij}} + \mathbf{F}_i^{(s)},
\end{align}
with $r_{ij}=\|\mathbf{r}_i-\mathbf{r}_j\|_2$ and $\mathbf{u}_{ij}=\mathbf{u}_i-\mathbf{u}_j$. Note that despite the absence of large reference density fractions, we include the smoothing effect of the inter-particle averaged pressure term.

To determine the volumetric surface force acting on each particle, we consider two different approaches: (1) based on the present PCP method and (2), an approach based on a colorfield function and SPH operators. For estimating mean curvature and surface normals using PCP, we use a fourth-order minter basis and threshold parameters of $\varepsilon=10^{-12}$ and $k_{\mathrm{max}}=1000$, a cutoff radius of $r_c=2.8h$, a sample-particle threshold $\xi=1.5h$, a narrow-band width of $w=15h$, and a skin width of $3w$.

To compute the interfacial surface tension, we require a smooth surface representation that distributes the surface tension effect on particles surrounding the interface. For this, we use the Wendland C2 kernel in 1D,
\begin{equation}
    W_{\epsilon1D}\left(q\right)=\begin{cases}
    \frac{5}{8\epsilon}\left(1-\frac{q}{2}\right)^3\left(1.5q+1\right)\quad &\text{if }0\leq q<2,\\
    0&\text{else},
    \end{cases}
\end{equation}
in which $q=\frac{|\phi|}{\epsilon}$. The smoothing length can in principle be chosen independently from the rest of the SPH operators, but we choose them to be identical for convenience. To detect particles relevant for the surface tension, we tune the skin redistancing frequency as described in the main text: For the three different resolutions considered, $h=1/32$, $h=1/64$, and $h=1/128$, we use $f=100, f=200$, and $f=400$, respectively. These frequencies do not only ensure that Eq.~\eqref{eq:skinfreq} is fulfilled, but also that particles relevant to the continuum surface dynamics are kept track of. For this, the cutoff radius of the SPH operators, $2\epsilon=6h$, needs to be considered instead of the extent of the regression domain, $r_c+\xi=2.8h+1.5h=4.3h$. The resulting skin redistanding frequencies are then determined from the maximum velocity $\|\mathbf{u}_{\mathrm{max}}\|_2=3$ found in the simulation. 

With the level-set SDF, the surface normals, and mean curvatures, the discrete surface force on particle $i$ is computed as
\begin{equation}\label{eq:pcpsurfaceforce}
    \mathbf{F}_i^{(s)}=-\frac{\tau}{\rho_i}\kappa_i\mathbf{n}_iW_{\epsilon1D}\left(\phi_i\right)\, .
\end{equation}

In the main text, we compare this with the popular colorfield-SPH approach \cite{morris2000simulating}.
Colorfield-SPH identifies the transition region of the interface and computes normals and curvatures from a {\it color function} $c$ assigning a unique color to each of the fluid phases. Following \cite{williams1998accuracy, morris2000simulating}, a smooth color value of a particle is obtained as a convolution, or particle function representation, of the binary indicator field,
\begin{equation}
    c_i=\sum_jc_j^{(ab)}W_{ij}V_j,
\end{equation}
where $c_j^{(ab)}$ is the binary phase indicator taking values of either 1 on one side of the interface or 0 on the other side, $W_{ij}$ here is the Wendland C2 kernel, and the volumes $V_j$ are computed according to Eq.~\eqref{eq:particlevolume}.
The remainder of the colorfield-based geometric computing framework is as outlined in Ref.~\cite{morris2000simulating}, where non-unit surface normals $\hat{\mathbf{n}}$ are obtained as
\begin{equation}
    \mathbf{\hat{n}}_i=\sum_j\left(c_j-c_i\right)\nabla W_{ij},
\end{equation}
whose magnitude is used as a indicator whether a particle is part of the smoothed interface or not, based on
\begin{equation}
    N_i=\begin{cases}
    1 \quad\text{if } \|\hat{\mathbf{n}}_i\|_2>0.01/\epsilon\\
    0 \quad\text{else.}
    \end{cases}
\end{equation}
Subsequently, narrow-banded unit normals can be obtained as
\begin{equation}
    \mathbf{n}_i=\begin{cases}
    \hat{\mathbf{n}}_i/\|\hat{\mathbf{n}}_i\|_2 \quad&\text{if }N_i=1,\\
    0&\text{else.}
    \end{cases}
\end{equation}
Finally, mean curvature is approximated by the SPH divergence of the unit surface normals from all ``interface particles'':
\begin{equation}
    \kappa_i=\left(\sum_j\text{min}\left(N_i,N_j\right)W_{ij}V_j\right)^{\!-1}\!\!\sum_j\text{min}\left(N_i,N_j\right)\left(\mathbf{n}_j-\mathbf{n}_i\right)\cdot\nabla W_{ij}V_j\, ,
\end{equation}
where the pre-factor accounts for differing contributions due to particles qualifying as interface particles, or not. In Eq.~\eqref{eq:csfmodel}, the gradient of the smoothed color function, $\hat{\mathbf{n}}_i$, is interpreted as the product of a regularized delta function and the surface normal, yielding the discrete surface force on particle $i$ as computed using colorfield-SPH:
\begin{equation}\label{eq:sphsurfaceforce}
    \mathbf{F}_i^{(s)}=-\frac{\tau}{\rho_i}\kappa_i\hat{\mathbf{n}}_i\, .
\end{equation}

Regardless of the approach chosen to determine the interfacial forces, PCP or colorfield-SPH, we integrate the positions of the particles in time with a second-order predictor-corrector scheme as in Ref.~\cite{zago2018semi} and apply the geometric computations at every predictor and every corrector step. We determine the time-step size $\Delta t$ such that it fulfills the CFL-like conditions given in Refs.~\cite{monaghan2005smoothed, adami2010new, morris2000simulating}:
\begin{equation}
    \Delta t\,\leq\,\min\left(0.25\frac{\epsilon}{c+\|\mathbf{u}_{\mathrm{max}}\|_2},~0.125\frac{\rho\epsilon^2}{\eta},~0.25\sqrt{\frac{\rho_0\epsilon^3}{2\pi\tau}}\right),
\end{equation}
where we again use $\|\mathbf{u}_{\mathrm{max}}\|_2=3$.

\section{Appendix figures}\label{sec:appfigures}

We provide additional figures as referred to and discussed in the main text.
\begin{figure}[ht]
    \centering
    \includegraphics{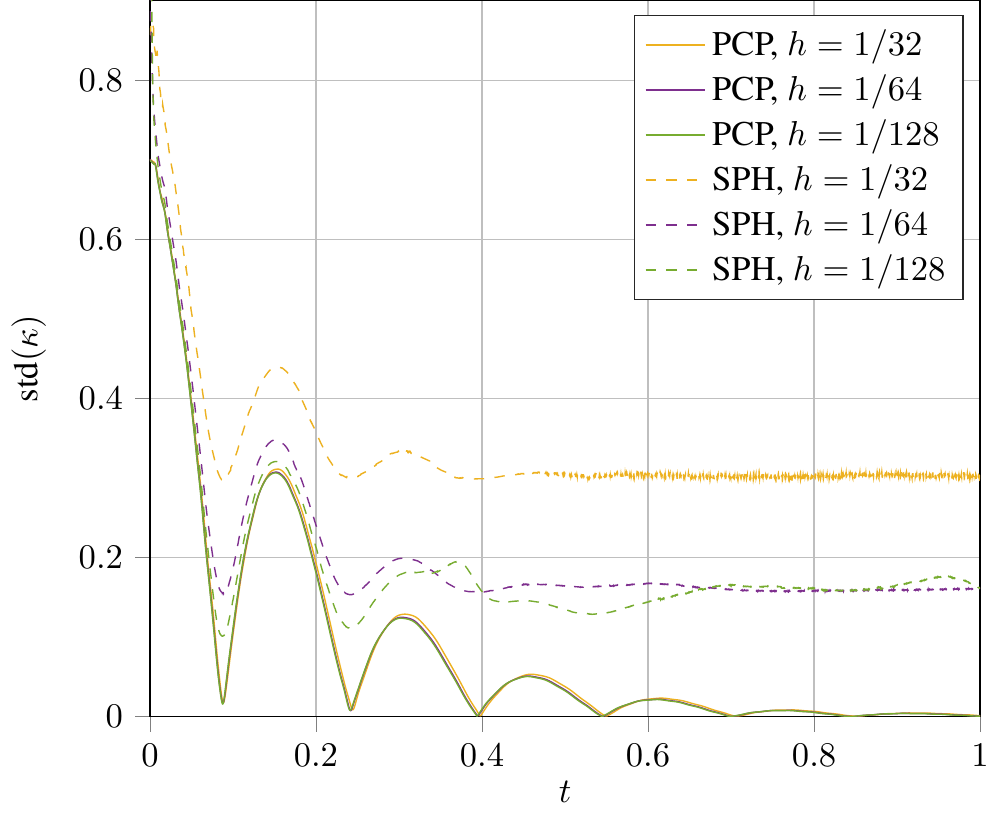}
    \caption{Standard deviation of the computed mean curvature values over the particles participating in continuum surface force approximation for different resolutions and methods.}
    \label{fig:stdkappa}
\end{figure}

\begin{figure}[ht]
    \centering
    \includegraphics{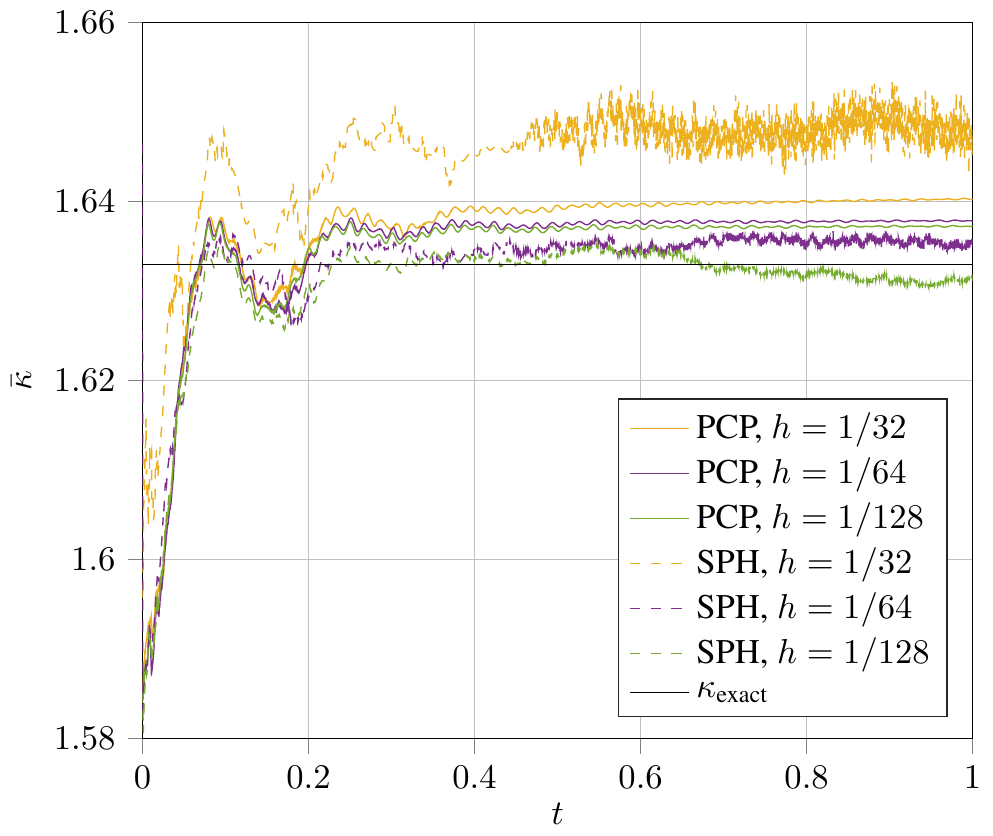}
    \caption{Mean of the computed mean curvature values over the particles participating in continuum surface force approximation for different resolutions and methods.}
    \label{fig:meankappa}
\end{figure}

\end{document}